\newcommand{\beq}{\begin{equation}}
\newcommand{\eeq}{\end{equation}}
\newcommand{\beqa}{\begin{eqnarray}}
\newcommand{\eeqa}{\end{eqnarray}}
\newcommand{\no}{\nonumber}
\def\lsim{\mathrel{\rlap{\lower4pt\hbox{\hskip1pt$\sim$}}
    \raise1pt\hbox{$<$}}}         
\def\gsim{\mathrel{\rlap{\lower4pt\hbox{\hskip1pt$\sim$}}
    \raise1pt\hbox{$>$}}}         
\newcommand{\re}[1]{\ensuremath{{\cal R}e(#1)}}
\newcommand{\im}[1]{\ensuremath{{\cal I}m(#1)}}
\newcommand{\Bz}{{B^0}}
\newcommand{\Bzb}{\overline{B}{}^0}

\newcommand{\CP}{CP\ }

\newcommand{\Bbar}{\overline{B}}

\newcommand{\Heff}{{\cal H}}
\newcommand{\Meff}{M}
\newcommand{\Geff}{\Gamma}

\newcommand{\fb}{\overline{f}}
\newcommand{\f}{f}    
    
\documentclass{cernrep}
\begin{document}
\title{Flavour Physics and CP Violation}
\author{Y. Nir 
}
\institute{Department of Particle Physics and Astrophysics\\
  Weizmann Institute of Science, Israel}
\maketitle

\begin{abstract}
We explain the many reasons for the interest in flavor physics. We describe flavor physics and the related CP violation within the Standard Model, and explain how the B-factories proved that the Kobayashi-Maskawa mechanism dominates the CP violation that is observed in meson decays. We explain the implications of flavor physics for new physics, with emphasis on the ``new physics flavor puzzle'', and present the idea of minimal flavor violation as a possible solution. We explain why the values flavor parameters of the Standard Model are puzzling, present the Froggatt-Nielsen mechanism as a possible solution, and describe how measurements of neutrino parameters are interpreted in the context of this puzzle. We show that the recently discovered Higgs-like boson may provide new opportunities for making progress on the various flavor puzzles.
\end{abstract}

\section{What is flavor?}
\label{sec:intro}
The term ``{\bf flavors}'' is used, in the jargon of particle physics, to
describe several copies of the same gauge representation, namely
several fields that are assigned the same quantum charges. Within the
Standard Model, when thinking of its unbroken $SU(3)_{\rm C}\times U(1)_{\rm
  EM}$ gauge group, there are four different types of particles, each
coming in three flavors:
\begin{itemize}
\item Up-type quarks in the $(3)_{+2/3}$ representation: $u,c,t$;
\item Down-type quarks in the $(3)_{-1/3}$ representation: $d,s,b$;
\item Charged leptons in the $(1)_{-1}$ representation: $e,\mu,\tau$;
\item Neutrinos in the $(1)_{0}$ representation: $\nu_1,\nu_2,\nu_3$.
\end{itemize}

The term ``{\bf flavor physics}'' refers to interactions that distinguish
between flavors. By definition, gauge interactions, namely
interactions that are related to unbroken symmetries and mediated
therefore by massless gauge bosons, do not distinguish among the
flavors and do not constitute part of flavor physics. Within the
Standard Model, flavor-physics refers to the weak and Yukawa
interactions.

The term ``{\bf flavor parameters}'' refers to parameters that carry
flavor indices. Within the Standard Model, these are the nine masses of
the charged fermions and the four ``mixing parameters'' (three angles
and one phase) that describe the interactions of the
charged weak-force carriers ($W^\pm$) with quark-antiquark pairs. If
one augments the Standard Model with Majorana mass terms for the
neutrinos, one should add to the list three neutrino masses and six
mixing parameters (three angles and three phases) for the $W^\pm$
interactions with lepton-antilepton pairs.

The term ``{\bf flavor universal}'' refers to interactions with couplings
(or to parameters) that are proportional to the unit matrix in
flavor space. Thus, the strong and electromagnetic interactions are
flavor-universal. An alternative term for ``flavor-universal'' is ``{\bf flavor-blind}''.

The term ``{\bf flavor diagonal}'' refers to interactions with couplings (or
to parameters) that are diagonal, but not necessarily
universal, in the flavor space. Within the Standard Model, the Yukawa
interactions of the Higgs particle are flavor diagonal.

The term ``{\bf flavor changing}'' refers to processes where the
initial and final flavor-numbers (that is, the number of particles of a
certain flavor minus the number of anti-particles of the same flavor)
are different. In ``flavor changing charged current'' processes, both
up-type and down-type flavors, and/or both charged lepton and neutrino
flavors are involved. Examples are (i) muon decay via $\mu\to
e\bar\nu_i\nu_j$, and (ii) $K^-\to\mu^-\bar\nu_j$ (which corresponds,
at the quark level, to $s\bar u\to\mu^-\bar\nu_j$). Within the
Standard Model, these processes are mediated by the $W$-bosons and
occur at tree level. In ``{\bf flavor changing neutral current}'' (FCNC)
processes, either up-type or down-type flavors but not both, and/or
either charged lepton or neutrino flavors but not both, are involved.
Example are (i) muon decay via $\mu\to e\gamma$ and (ii)
$K_L\to\mu^+\mu^-$ (which corresponds, at the quark level, to $s\bar
d\to\mu^+\mu^-$). Within the Standard Model, these processes do not
occur at tree level, and are often highly suppressed.

Another useful term is ``{\bf flavor violation}''. We explain it later
in these lectures.

\section{Why is flavor physics interesting?}
\label{sec:mot}
\begin{itemize}
\item Flavor physics can discover new physics or probe it before it is
  directly observed in experiments. Here are some examples from the
  past:
\begin{itemize}
\item The smallness of $\frac{\Gamma(K_L\to\mu^+\mu^-)}
  {\Gamma(K^+\to\mu^+\nu)}$ led to predicting a fourth (the charm)
  quark;
\item The size of $\Delta m_K$ led to a successful prediction of the
  charm mass;
\item The size of $\Delta m_B$ led to a successful prediction of the
  top mass;
\item The measurement of $\varepsilon_K$ led to predicting the third
  generation.
\item The measurement of neutrino flavor transitions led to the
  discovery of neutrino masses.
\end{itemize}
\item CP violation is closely related to flavor physics. Within the
  Standard Model, there is a single CP violating parameter, the
  Kobayashi-Maskawa phase $\delta_{\rm KM}$ \cite{Kobayashi:1973fv}.
  Baryogenesis tells us, however, that there must exist new sources of
  CP violation. Measurements of CP violation in flavor changing
  processes might provide evidence for such sources.
\item The fine-tuning problem of the Higgs mass, and the puzzle of the
  dark matter imply that there exists new physics at, or below, the
  TeV scale. If such new physics had a generic flavor structure, it
  would contribute to flavor changing neutral current (FCNC) processes
  orders of magnitude above the observed rates. The question of why
  this does not happen constitutes the {\it new physics flavor
    puzzle}.
\item Most of the charged fermion flavor parameters are small and
  hierarchical. The Standard Model does not provide any explanation of
  these features. This is the {\it Standard Model flavor puzzle}.  The
  puzzle became even deeper after neutrino masses and mixings were
  measured because, so far, neither smallness nor hierarchy in these
  parameters have been established.
\end{itemize}

\section{Flavor in the Standard Model}
\label{smfor}
A model of elementary particles and their interactions is defined
by the following ingredients: (i) The symmetries of the Lagrangian and
the pattern of spontaneous symmetry breaking; (ii) The representations
of fermions and scalars. The Standard Model (SM) is defined as follows:\\
(i) The gauge symmetry is
\beq\label{smsym}
G_{\rm SM}=SU(3)_{\rm C}\times SU(2)_{\rm L}\times U(1)_{\rm Y}.
\eeq
It is spontaneously broken by the VEV of a single Higgs scalar,
$\phi(1,2)_{1/2}$  $\left(\langle\phi^0\rangle=v/\sqrt{2}\right)$:
\beq\label{smssb}
G_{\rm SM} \to SU(3)_{\rm C}\times U(1)_{\rm EM}.
\eeq
(ii) There are three fermion generations, each consisting of five
representations of $G_{\rm SM}$:
\beq\label{ferrep}
Q_{Li}(3,2)_{+1/6},\ \ U_{Ri}(3,1)_{+2/3},\ \
D_{Ri}(3,1)_{-1/3},\ \ L_{Li}(1,2)_{-1/2},\ \ E_{Ri}(1,1)_{-1}.
\eeq
%

\subsection{The interaction basis}
The Standard Model Lagrangian, ${\cal L}_{\rm SM}$, is the most
general renormalizable Lagrangian that is consistent with the gauge
symmetry (\ref{smsym}), the particle content (\ref{ferrep}) and the
pattern of spontaneous symmetry breaking (\ref{smssb}). It can be
divided to three parts:
\beq\label{LagSM}
{\cal L}_{\rm SM}={\cal L}_{\rm kinetic}+{\cal L}_{\rm Higgs}
+{\cal L}_{\rm Yukawa}.
\eeq

As concerns the kinetic terms, to maintain gauge invariance, one has
to replace the derivative with a covariant derivative:
\beq\label{SMDmu}
D^\mu=\partial^\mu+ig_s G^\mu_a L_a+ig W^\mu_b T_b+ig^\prime B^\mu Y.
\eeq
Here $G^\mu_a$ are the eight gluon fields, $W^\mu_b$ the three
weak interaction bosons and $B^\mu$ the single hypercharge boson.
The $L_a$'s are $SU(3)_{\rm C}$ generators (the $3\times3$
Gell-Mann matrices ${1\over2}\lambda_a$ for triplets, $0$ for singlets),
the $T_b$'s are $SU(2)_{\rm L}$ generators (the $2\times2$
Pauli matrices ${1\over2}\tau_b$ for doublets, $0$ for singlets),
and the $Y$'s are the $U(1)_{\rm Y}$ charges. For example, for the
quark doublets $Q_L$, we have
\beq\label{DmuQL}
{\cal L}_{\rm kinetic}(Q_L)= i{\overline{Q_{Li}}}\gamma_\mu
\left(\partial^\mu+{i\over2}g_s G^\mu_a\lambda_a
+{i\over2}g W^\mu_b\tau_b+{i\over6}g^\prime
B^\mu\right)\delta_{ij}Q_{Lj},
\eeq
while for the lepton doublets $L_L^I$, we have
\beq\label{DmuLL}
{\cal L}_{\rm kinetic}(L_L)= i{\overline{L_{Li}}}\gamma_\mu
\left(\partial^\mu+{i\over2}g W^\mu_b\tau_b-\frac i2 g^\prime
  B^\mu\right)\delta_{ij}L_{Lj}.
\eeq
The unit matrix in flavor space, $\delta_{ij}$, signifies that
these parts of the interaction Lagrangian are flavor-universal. In
addition, they conserve CP.

The Higgs potential, which describes the scalar self interactions, is given by:
\beq\label{HiPo}
{\cal L}_{\rm Higgs}=\mu^2\phi^\dagger\phi-\lambda(\phi^\dagger\phi)^2.
\eeq
For the Standard Model scalar sector, where there is a single doublet,
this part of the Lagrangian is also CP conserving.

The quark Yukawa interactions are given by
\beq\label{Hqint}
-{\cal L}_{\rm Y}^{q}=Y^d_{ij}{\overline {Q_{Li}}}\phi D_{Rj}
+Y^u_{ij}{\overline {Q_{Li}}}\tilde\phi U_{Rj}+{\rm h.c.},
\eeq
(where $\tilde\phi=i\tau_2\phi^\dagger$) while the lepton Yukawa
interactions are given by
\beq\label{Hlint}
-{\cal L}_{\rm Y}^{\ell}=Y^e_{ij}{\overline {L_{Li}}}\phi E_{Rj}
+{\rm h.c.}.
\eeq
This part of the Lagrangian is, in general, flavor-dependent (that is,
$Y^f\not\propto{\bf 1}$) and CP violating.

\subsection{Global symmetries}
\label{sec:spurions}
In the absence of the Yukawa matrices $Y^d$, $Y^u$ and $Y^e$, the SM
has a large $U(3)^5$ global symmetry:
\beq\label{gglobal}
G_{\rm global}(Y^{u,d,e}=0)=SU(3)_q^3\times SU(3)_\ell^2\times U(1)^5,
\eeq
where
\beqa\label{susuu}
SU(3)_q^3&=&SU(3)_Q\times SU(3)_U\times SU(3)_D,\no\\
SU(3)_\ell^2&=&SU(3)_L\times SU(3)_E,\no\\
U(1)^5&=&U(1)_B\times U(1)_L\times U(1)_Y\times U(1)_{\rm PQ}\times
U(1)_E.
\eeqa
Out of the five $U(1)$ charges, three can be identified with baryon
number ($B$), lepton number ($L$) and hypercharge ($Y$), which are
respected by the Yukawa interactions. The two remaining $U(1)$ groups
can be identified with the PQ symmetry whereby the Higgs and $D_R,E_R$
fields have opposite charges, and with a global rotation of $E_R$
only.

The point that is important for our purposes is that ${\cal L}_{\rm
  kinetic}+{\cal L}_{\rm Higgs}$ respect the non-Abelian flavor
symmetry $S(3)_q^3\times SU(3)_\ell^2$, under which
\beq\label{symkh}
Q_L\to V_QQ_L,\ \ \ U_R\to V_U U_R,\ \ \ D_R\to V_D D_R,\ \ L_L\to V_L
L_L,\ \ \ E_R\to V_E E_R,
\eeq
where the $V_i$ are unitary matrices.
The Yukawa interactions (\ref{Hqint}) and (\ref{Hlint}) break the
global symmetry,
\beq\label{globre}
G_{\rm global}(Y^{u,d,e}\neq0)= U(1)_B\times U(1)_e\times
U(1)_\mu\times U(1)_\tau.
\eeq
(Of course, the gauged $U(1)_Y$ also remains a good symmetry.)
Thus, the transformations of Eq. (\ref{symkh}) are not a symmetry of
${\cal L}_{\rm SM}$. Instead, they correspond to a change of the
interaction basis. These observations also offer an alternative way of
defining flavor physics: it refers to interactions that break the
$SU(3)^5$ symmetry (\ref{symkh}). Thus, the term ``{\bf flavor
  violation}'' is often used to describe processes or parameters that
break the symmetry.

One can think of the quark Yukawa couplings as spurions that break the
global $SU(3)_q^3$ symmetry (but are neutral under $U(1)_B$),
\beq\label{Gglobq}
Y^u\sim(3,\bar3,1)_{SU(3)_q^3},\ \ \
Y^d\sim(3,1,\bar3)_{SU(3)_q^3},
\eeq
and of the lepton Yukawa couplings as spurions that break the global
$SU(3)_\ell^2$ symmetry (but are neutral under $U(1)_e\times
U(1)_\mu\times U(1)_\tau$),
\beq\label{Gglobl}
Y^e\sim(3,\bar3)_{SU(3)_\ell^2}.
\eeq
The spurion formalism is convenient for several purposes: parameter
counting (see below), identification of flavor suppression factors
(see Section \ref{sec:nppuzzle}), and the idea of minimal flavor
violation (see Section \ref{sec:mfv}).

\subsection{Counting parameters}
How many independent parameters are there in ${\cal L}_{\rm Y}^q$? The
two Yukawa matrices, $Y^u$ and $Y^d$, are $3\times3$ and complex.
Consequently, there are 18 real and 18 imaginary parameters in these
matrices. Not all of them are, however, physical. The pattern of
$G_{\rm global}$ breaking means that there is freedom to remove 9 real
and 17 imaginary parameters (the number of parameters in three
$3\times3$ unitary matrices minus the phase related to $U(1)_B$). For
example, we can use the unitary transformations $Q_L\to V_QQ_L$,
$U_R\to V_U U_R$ and $D_R\to V_D D_R$, to lead to the following
interaction basis:
\beq\label{speint}
Y^d=\lambda_d,\ \ \ Y^u=V^\dagger\lambda_u,
\eeq
where $\lambda_{d,u}$ are diagonal,
\beq\label{deflamd}
\lambda_d={\rm diag}(y_d,y_s,y_b),\ \ \
\lambda_u={\rm diag}(y_u,y_c,y_t),
\eeq
while $V$ is a unitary matrix that depends on three real angles and
one complex phase. We conclude that there are 10 quark flavor
parameters: 9 real ones and a single phase. In the mass basis, we will
identify the nine real parameters as six quark masses and
three mixing angles, while the single phase is $\delta_{\rm KM}$.

How many independent parameters are there in ${\cal L}_{\rm Y}^\ell$?
The Yukawa matrix $Y^e$ is $3\times3$ and complex. Consequently, there
are 9 real and 9 imaginary parameters in this matrix. There is,
however, freedom to remove 6 real and 9 imaginary parameters (the
number of parameters in two $3\times3$ unitary matrices minus the
phases related to $U(1)^3$). For example, we can use the unitary
transformations $L_L\to V_LL_L$ and $E_R\to V_E E_R$, to lead to the
following interaction basis:
\beq\label{speintl}
Y^e=\lambda_e={\rm diag}(y_e,y_\mu,y_\tau).
\eeq
We conclude that there are 3 real lepton flavor parameters. In the
mass basis, we will identify these parameters as the three
charged lepton masses. We must, however, modify the model when
we take into account the evidence for neutrino masses.

\subsection{The mass basis}
Upon the replacement $\re{\phi^0}\to\frac{v+h^0}{\sqrt2}$, the Yukawa
interactions (\ref{Hqint}) give rise to the mass matrices
\beq\label{YtoMq}
M_q={v\over\sqrt2}Y^q.
\eeq
The mass basis corresponds, by definition, to diagonal mass
matrices. We can  always find unitary matrices $V_{qL}$ and $V_{qR}$
such that
\beq\label{diagMq}
V_{qL}M_q V_{qR}^\dagger=M_q^{\rm diag}\equiv\frac{v}{\sqrt2}\lambda_q.
\eeq
The four matrices $V_{dL}$, $V_{dR}$, $V_{uL}$ and $V_{uR}$ are then
the ones required to transform to the mass basis. For example, if we
start from the special basis (\ref{speint}), we have
$V_{dL}=V_{dR}=V_{uR}={\bf 1}$ and $V_{uL}=V$. The combination
$V_{uL}V_{dL}^\dagger$ is independent of the interaction basis from
which we start this procedure.

We denote the left-handed quark mass eigenstates as $U_L$ and $D_L$.
The charged current interactions for quarks [that is the interactions of the
charged $SU(2)_{\rm L}$ gauge bosons $W^\pm_\mu={1\over\sqrt2}
(W^1_\mu\mp iW_\mu^2)$], which in the interaction basis are described
by (\ref{DmuQL}), have a complicated form in the mass basis:
\beq\label{Wmasq}
-{\cal L}_{W^\pm}^q={g\over\sqrt2}{\overline {U_{Li}}}\gamma^\mu
V_{ij}D_{Lj} W_\mu^++{\rm h.c.}.
\eeq
where $V$ is the $3\times3$ unitary matrix ($VV^\dagger=V^\dagger
V={\bf 1}$) that appeared in Eq. (\ref{speint}). For a general
interaction basis,
\beq\label{VCKM}
V=V_{uL}V_{dL}^\dagger.
\eeq
$V$ is the Cabibbo-Kobayashi-Maskawa (CKM) {\it mixing matrix} for
quarks \cite{Cabibbo:1963yz,Kobayashi:1973fv}. As a result of the fact
that $V$ is not diagonal, the $W^\pm$ gauge bosons couple to quark
mass eigenstates of different generations. Within the Standard
Model, this is the only source of {\it flavor changing} quark
interactions.

{\bf Exercise 1:} {\it Prove that, in the absence of neutrino masses, there
is no mixing in the lepton sector.}

{\bf Exercise 2:} {\it Prove that there is no mixing in the $Z$
couplings. (In the physics jargon, there are no flavor changing
neutral currents at tree level.)}

The detailed structure of the CKM matrix, its parametrization, and the
constraints on its elements are described in Appendix \ref{app:ckm}.

\section{Testing CKM}
\label{sec:bmix}
Measurements of rates, mixing, and CP asymmetries in $B$ decays in the
two B factories, BaBar abd Belle, and in the two Tevatron detectors,
CDF and D0, signified a new era in our understanding of CP
violation. The progress is both qualitative and quantitative. Various
basic questions concerning CP and flavor violation have received, for
the first time, answers based on experimental information. These
questions include, for example,
\begin{itemize}
\item Is the Kobayashi-Maskawa mechanism at work (namely, is
  $\delta_{\rm KM}\neq0$)?
\item Does the KM phase dominate the observed CP violation?
\end{itemize}
As a first step, one may assume the SM and test the overall
consistency of the various measurements. However, the richness of data
from the B factories allow us to go a step further and answer these
questions model independently, namely allowing new physics to
contribute to the relevant processes. We here explain the way in which
this analysis proceeds.

\subsection{$S_{\psi K_S}$}
The CP asymmetry in $B\to\psi K_S$ decays plays a major role in
testing the KM mechanism. Before we explain the test itself, we should
understand why the theoretical interpretation of the asymmetry is
exceptionally clean, and what are the theoretical parameters on which
it depends, within and beyond the Standard Model.

The CP asymmetry in neutral meson decays into final CP eigenstates
$f_{\CP}$ is defined as follows:
\beq\label{asyfcpt}
{\cal A}_{f_{\CP}}(t)\equiv\frac{d\Gamma/dt[\Bzb_{\rm phys}(t)\to f_{\CP}]-
d\Gamma/dt[\Bz_{\rm phys}(t)\to f_{\CP}]}
{d\Gamma/dt[\Bzb_{\rm phys}(t)\to f_{\CP}]+d\Gamma/dt[\Bz_{\rm phys}(t)\to
  f_{\CP}]}\; .
\eeq
A detailed evaluation of this asymmetry is given in Appendix
\ref{sec:formalism}. It leads to the following form:
\beqa\label{asyfcpbt}
{\cal A}_{f_{\CP}}(t)&=&S_{f_{\CP}}\sin(\Delta
mt)-C_{f_{\CP}}\cos(\Delta mt),\no\\
S_{f_{\CP}}&\equiv&\frac{2\,\im{\lambda_{f_{\CP}}}}{1+|\lambda_{f_{\CP}}|^2},\
\ \  C_{f_{\CP}}\equiv\frac{1-|\lambda_{f_{\CP}}|^2}{1+|\lambda_{f_{\CP}}|^2}
\; ,
\eeqa
where
\beq\label{lamhad}
\lambda_{f_{\CP}}=e^{-i\phi_B}(\overline{A}_{f_{\CP}}/A_{f_{\CP}}) \; .
\eeq
Here $\phi_B$ refers to the phase of $M_{12}$ [see
Eq.~(\ref{defmgam})].  Within the Standard Model, the corresponding
phase factor is given by
\beq\label{phimsm}
e^{-i\phi_B}=(V_{tb}^* V_{td}^{})/(V_{tb}^{}V_{td}^*) \;.
\eeq
The decay amplitudes $A_f$ and $\overline{A}_f$ are defined in
Eq. (\ref{decamp}).

\begin{figure}[htb]
\begin{center}
\includegraphics[width=2.85in]{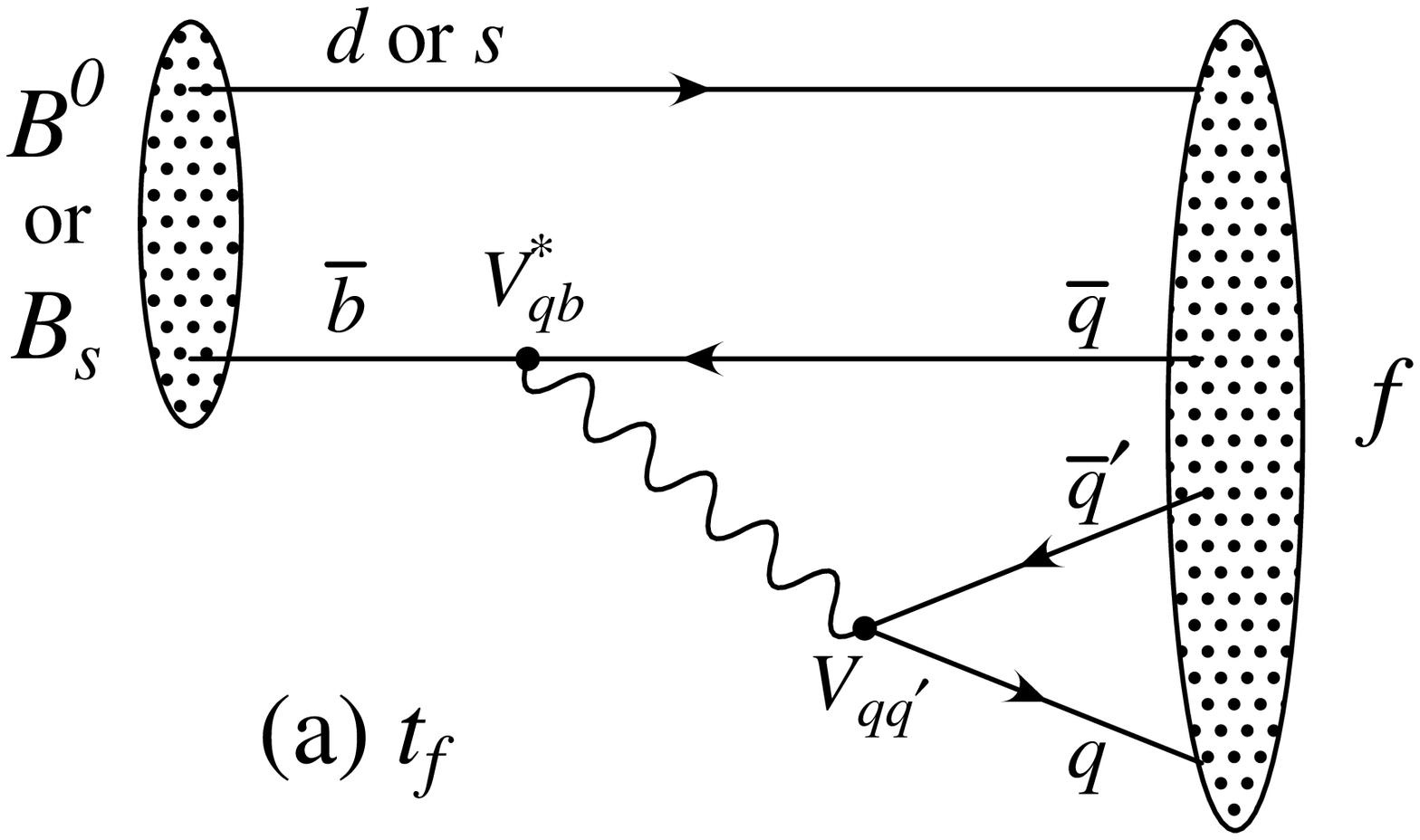}
\hspace{2em}
\includegraphics[width=2.85in]{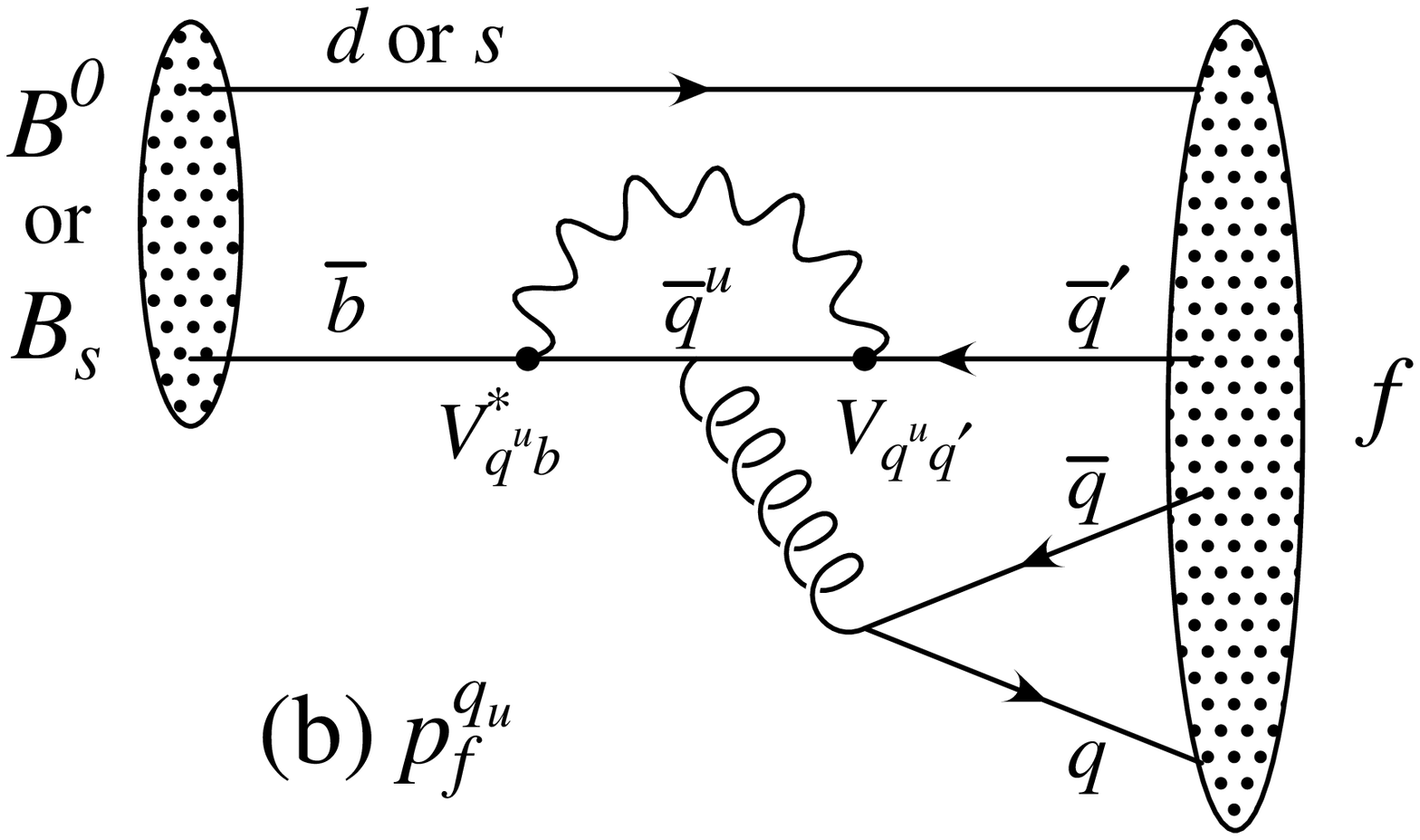}
\caption{Feynman diagrams for (a) tree and (b) penguin amplitudes
  contributing to $B^0\to f$ or $B_{s}\to f$ via a $\bar b\to\bar q
  q\bar q^\prime$ quark-level process.}
\label{fig:diags}
\end{center}
\end{figure}

The $B^0\to J/\psi K^0$ decay~\cite{Carter:1980hr,Bigi:1981qs} proceeds
via the quark transition $\bar b\to\bar c c\bar s$. There are
contributions from both tree ($t$) and penguin ($p^{q_u}$, where
$q_u=u,c,t$ is the quark in the loop) diagrams (see
Fig.~\ref{fig:diags}) which carry different weak phases:
\beq\label{ckmdec}
A_f = \left(V^\ast_{cb} V^{}_{cs}\right) t_f +
\sum_{q_u= u,c,t}\left(V^\ast_{q_u b} V^{}_{q_u s}\right) p^{q_u}_f \; .
\eeq
(The distinction between tree and penguin contributions is a heuristic
one, the separation by the operator that enters is more precise. For a
detailed discussion of the more complete operator product approach,
which also includes higher order QCD corrections, see, for example,
ref. \cite{Buchalla:1995vs}.)  Using CKM unitarity, these decay
amplitudes can always be written in terms of just two CKM
combinations:
\beq\label{btoccs}
A_{\psi K}=\left(V^\ast_{cb} V^{}_{cs}\right)T_{\psi
  K}+\left(V^\ast_{ub} V^{}_{us}\right)P^u_{\psi K},
\eeq
where $T_{\psi K}=t_{\psi K}+p^c_{\psi K}-p^t_{\psi K}$ and
$P^u_{\psi K}=p^u_{\psi K}-p^t_{\psi K}$. A subtlety arises in this
decay that is related to the fact that ${B}^0\to J/\psi K^0$ and
$\overline{B}^0\to J/\psi\overline{K}{}^0$. A common final state,
e.g. $J/\psi K_S$, can be reached via $K^0-\overline{K}{}^0$ mixing.
Consequently, the phase factor corresponding to neutral $K$ mixing,
$e^{-i\phi_K}=(V^*_{cd}V^{}_{cs})/(V^{}_{cd}V^*_{cs})$, plays a
role:
\beq\label{psikmix}
\frac{\overline{A}_{\psi K_S}}{A_{\psi K_S}}
=-\frac{\left(V^{}_{cb} V^\ast_{cs}\right)T_{\psi
    K}+\left(V^{}_{ub} V^\ast_{us}\right)P^u_{\psi K}}
{\left(V^\ast_{cb} V^{}_{cs}\right)T_{\psi
    K}+\left(V^\ast_{ub} V^{}_{us}\right)P^u_{\psi K}}\times
\frac{V_{cd}^\ast V_{cs}^{}}{V_{cd}^{}V_{cs}^\ast}.
\eeq

The crucial point is that, for $B\to J/\psi K_S$ and other $\bar
b\to\bar cc\bar s$ processes, we can neglect the $P^u$ contribution to
$A_{\psi K}$, in the SM, to an approximation that is better than one
percent:
\beq\label{smapprox}
|P^u_{\psi K}/T_{\psi K}|\times|V_{ub}/V_{cb}|\times|
V_{us}/V_{cs}|\sim({\rm loop\ factor})\times0.1\times0.23\lsim0.005.
\eeq
Thus, to an accuracy better than one percent,
\beq
\lambda_{\psi K_S}=\left(\frac{V_{tb}^*
  V_{td}^{}}{V_{tb}^{}V_{td}^*}\right)\left(\frac{V_{cb}
  V_{cd}^{*}}{V_{cb}^{*}V_{cd}}\right)=-e^{-2i\beta},
\eeq
where $\beta$ is defined in Eq. (\ref{abcangles}), and consequently
\beq\label{btopsik}
S_{\psi K_S}=\sin2\beta,\ \ \ C_{\psi K_S}=0 \; .
\eeq
(Below the percent level, several effects modify this equation
\cite{Grossman:2002bu,Boos:2004xp,Li:2006vq,Gronau:2008cc}.)

{\bf Exercise 3:} {\it Show that, if the $B\to\pi\pi$ decays were dominated
by tree diagrams, then $S_{\pi\pi}=\sin2\alpha$.}

{\bf Exercise 4:} {\it Estimate the accuracy of the predictions
$S_{\phi K_S}=\sin2\beta$ and $C_{\phi K_S}=0$.}

When we consider extensions of the SM, we still do not expect any
significant new contribution to the tree level decay, $b\to c\bar cs$,
beyond the SM $W$-mediated diagram. Thus, the expression $\bar A_{\psi
  K_S}/A_{\psi K_S}=(V_{cb}V_{cd}^*)/(V_{cb}^*V_{cd})$ remains valid,
though the approximation of neglecting sub-dominant phases can be
somewhat less accurate than Eq. (\ref{smapprox}). On the other hand,
$M_{12}$, the $B^0-\overline{B}^0$ mixing amplitude, can in principle
get large and even dominant contributions from new physics. We can
parametrize the modification to the SM in terms of two parameters,
$r_d^2$ signifying the change in magnitude, and $2\theta_d$
signifying the change in phase:
\beq\label{derthed}
M_{12}=r_d^2\ e^{2i\theta_d}\ M_{12}^{\rm SM}(\rho,\eta).
\eeq
This leads to the following generalization of Eq. (\ref{btopsik}):
\beq\label{btopsiknp}
S_{\psi K_S}=\sin(2\beta+2\theta_d),\ \ \ C_{\psi K_S}=0 \; .
\eeq

The experimental measurements give the following ranges
\cite{Amhis:2012bh}:
\beq\label{scpkexp}
S_{\psi K_S}=+0.68\pm0.02,\ \ \ C_{\psi K_S}=+0.005\pm0.017 \; .
\eeq

\subsection{Self-consistency of the CKM assumption}
The three generation standard model has room for CP violation, through
the KM phase in the quark mixing matrix. Yet, one would like to make
sure that indeed CP is violated by the SM interactions, namely that
$\sin\delta_{\rm KM}\neq0$. If we establish that this is the case, we
would further like to know whether the SM contributions to CP
violating observables are dominant. More quantitatively, we would like
to put an upper bound on the ratio between the new physics and the SM
contributions.

\begin{figure}[tb]
  \centering
  {\includegraphics[width=0.65\textwidth]{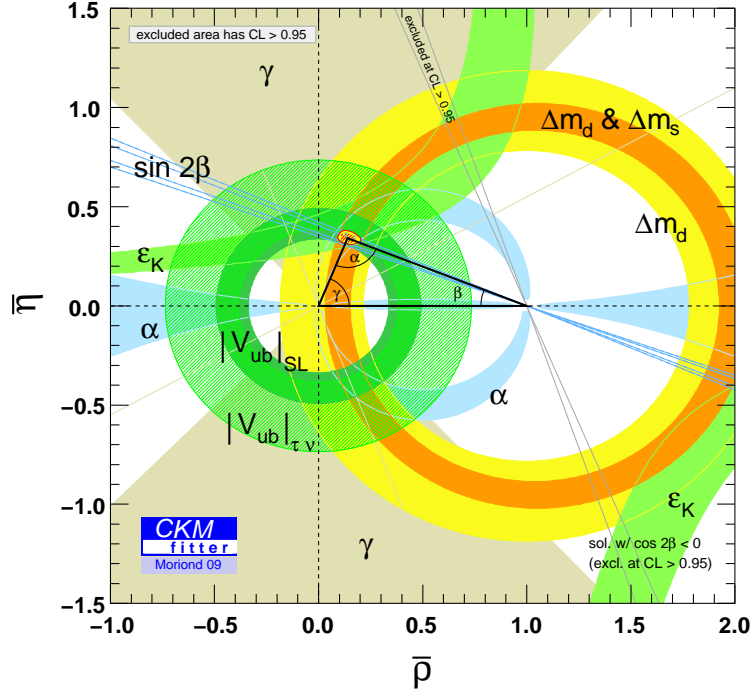}}
  \caption{Allowed region in the $\rho,\eta$ plane. Superimposed are
  the individual constraints from charmless semileptonic $B$ decays
  ($|V_{ub}/V_{cb}|$), mass differences in the $B^0$ ($\Delta m_d$)
  and $B_s$ ($\Delta m_s$) neutral meson systems, and CP violation in
  $K\to\pi\pi$ ($\varepsilon_K$), $B\to\psi K$ ($\sin2\beta$),
  $B\to\pi\pi,\rho\pi,\rho\rho$ ($\alpha$), and $B\to DK$
  ($\gamma$). Taken from \cite{ckmfitter}.}
  \label{fg:UT}
\end{figure}

As a first step, one can assume that flavor changing processes are
fully described by the SM, and check the consistency of the various
measurements with this assumption. There are four relevant mixing
parameters, which can be taken to be the Wolfenstein parameters
$\lambda$, $A$, $\rho$ and $\eta$ defined in Eq. (\ref{wolpar}). The
values of $\lambda$ and $A$ are known rather accurately
\cite{Beringer:1900zz} from, respectively, $K\to\pi\ell\nu$ and $b\to
c\ell\nu$ decays:
\beq\label{lamaexp}
\lambda=0.2254\pm0.0007,\ \ \ A=0.811^{+0.022}_{-0.012}.
\eeq
Then, one can express all the relevant observables as a function of
the two remaining parameters, $\rho$ and $\eta$, and check whether
there is a range in the $\rho-\eta$ plane that is consistent with all
measurements. The list of observables includes the following:
\begin{itemize}
\item The rates of inclusive and exclusive charmless semileptonic $B$
  decays depend on $|V_{ub}|^2\propto\rho^2+\eta^2$;
\item The CP asymmetry in $B\to\psi K_S$, $S_{\psi
    K_S}=\sin2\beta=\frac{2\eta(1-\rho)}{(1-\rho)^2+\eta^2}$;
\item The rates of various $B\to DK$ decays depend on the phase
  $\gamma$, where $e^{i\gamma}=\frac{\rho+i\eta}{\sqrt{\rho^2+\eta^2}}$;
\item The rates of various $B\to\pi\pi,\rho\pi,\rho\rho$ decays depend
  on the phase $\alpha=\pi-\beta-\gamma$;
\item The ratio between the mass splittings in the neutral $B$ and
  $B_s$ systems is sensitive to $|V_{td}/V_{ts}|^2=\lambda^2[(1-\rho)^2+\eta^2]$;
\item The CP violation in $K\to\pi\pi$ decays, $\epsilon_K$, depends
  in a complicated way on $\rho$ and $\eta$.
\end{itemize}
The resulting constraints are shown in Fig. \ref{fg:UT}.

The consistency of the various constraints is impressive. In
particular, the following ranges for $\rho$ and $\eta$ can account for
all the measurements \cite{Beringer:1900zz}:
\beq
\rho=+0.131^{+0.026}_{-0.013},\ \ \ \eta=+0.345\pm0.014.
\eeq

One can make then the following statement \cite{Nir:2002gu}:\\
{\bf Very likely, CP violation in flavor changing processes is
  dominated by the Kobayashi-Maskawa phase.}

In the next two subsections, we explain how we can remove the phrase
``very likely'' from this statement, and how we can quantify the
KM-dominance.

\subsection{Is the KM mechanism at work?}
In proving that the KM mechanism is at work, we assume that
charged-current tree-level processes are dominated by the $W$-mediated
SM diagrams (see, for example, \cite{Grossman:1997dd}).
This is a very plausible assumption. I am not aware of
any viable well-motivated model where this assumption is not valid.
Thus we can use all tree level processes and fit them to $\rho$ and
$\eta$, as we did before. The list of such processes includes the
following:
\begin{enumerate}
\item Charmless semileptonic $B$-decays, $b\to u\ell\nu$, measure
  $R_u$ [see Eq. (\ref{RbRt})].
\item $B\to DK$ decays, which go through the quark transitions $b\to
  c\bar u s$ and $b\to u\bar cs$, measure the angle $\gamma$ [see Eq.
  (\ref{abcangles})].
\item $B\to\rho\rho$ decays (and, similarly, $B\to\pi\pi$ and
  $B\to\rho\pi$ decays) go through the quark transition $b\to u\bar
  ud$. With an isospin analysis, one can determine the relative phase
  between the tree decay amplitude and the mixing amplitude. By
  incorporating the measurement of $S_{\psi K_S}$, one can subtract
  the phase from the mixing amplitude, finally providing a measurement
  of the angle $\gamma$ [see Eq. (\ref{abcangles})].
  \end{enumerate}

In addition, we can use loop processes, but then we must allow for new
physics contributions, in addition to the $(\rho,\eta)$-dependent SM
contributions. Of course, if each such measurement adds a separate
mode-dependent parameter, then we do not gain anything by using this
information. However, there is a number of observables where the only
relevant loop process is $B^0-\overline{B}{}^0$ mixing. The list
includes $S_{\psi K_S}$, $\Delta m_B$ and the CP asymmetry in
semileptonic $B$ decays:
\beqa\label{apksNP}
S_{\psi K_S}&=&\sin(2\beta+2\theta_d),\no\\
\Delta m_{B}&=&r_d^2(\Delta m_B)^{\rm SM},\no\\
{\cal A}_{\rm SL}&=&-{\cal
    R}e\left(\frac{\Gamma_{12}}{M_{12}}\right)^{\rm
    SM}\frac{\sin2\theta_d}{r_d^2}
  +{\cal I}m\left(\frac{\Gamma_{12}}{M_{12}}\right)^{\rm
    SM}\frac{\cos2\theta_d}{r_d^2}.
\eeqa
As explained above, such processes involve two new parameters [see Eq.
(\ref{derthed})]. Since there are three relevant observables, we can
further tighten the constraints in the $(\rho,\eta)$-plane. Similarly,
one can use measurements related to $B_s-\overline{B}_s$ mixing. One
gains three new observables at the cost of two new parameters (see,
for example, \cite{Grossman:2006ce}).

\begin{figure}[tb]
  \centering
  {\includegraphics[width=0.65\textwidth]{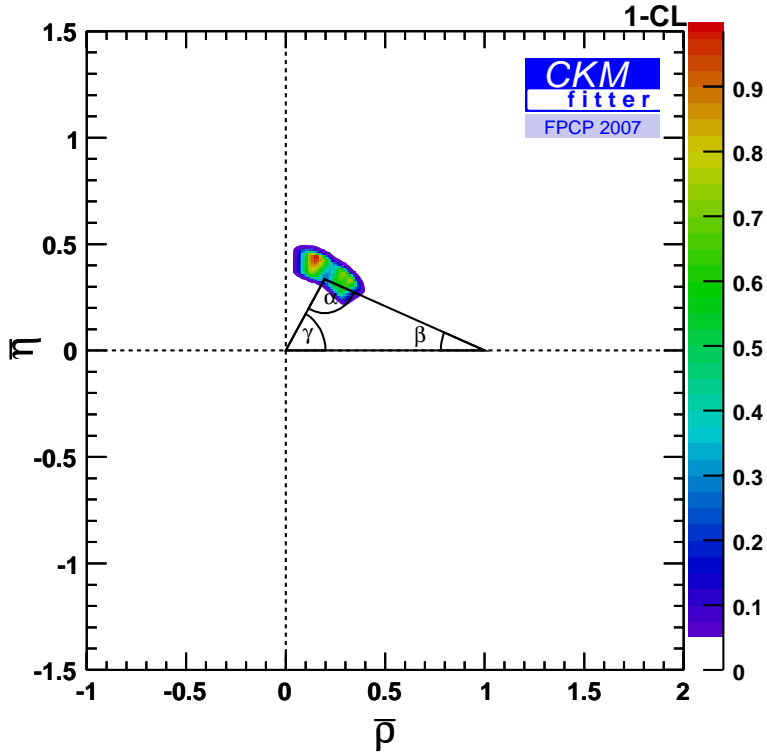}}
  \caption{The allowed region in the $\rho-\eta$ plane, assuming that
  tree diagrams are dominated by the Standard Model \cite{ckmfitter}.}
  \label{fig:re_tree}
\end{figure}

The results of such fit, projected on the $\rho-\eta$ plane, can be
seen in Fig. \ref{fig:re_tree}. It gives \cite{ckmfitter}
\beq
\eta=0.44^{+0.05}_{-0.23}\ \ (3\sigma).
\eeq
[A similar analysis in Ref. \cite{Bona:2007vi} obtains the $3\sigma$
range $(0.31-0.46)$.] It is clear that $\eta\neq0$ is well
established:\\
{\bf The Kobayashi-Maskawa mechanism of CP violation is at work.}

Another way to establish that CP is violated by the CKM matrix is to
find, within the same procedure, the allowed range for $\sin2\beta$
\cite{Bona:2007vi}:
\beq\label{stbth}
\sin2\beta^{\rm tree}=0.80\pm0.03.
\eeq
Thus, $\beta\neq0$ is well established.

The consistency of the experimental results (\ref{scpkexp}) with the
SM predictions (\ref{btopsik},\ref{stbth}) means that the KM mechanism
of CP violation dominates the observed CP violation. In the next
subsection, we make this statement more quantitative.

\subsection{How much can new physics contribute to
  $B^0-\overline{B}{}^0$ mixing?}
\begin{figure}[htb]
\begin{center}
\includegraphics[width=2.85in]{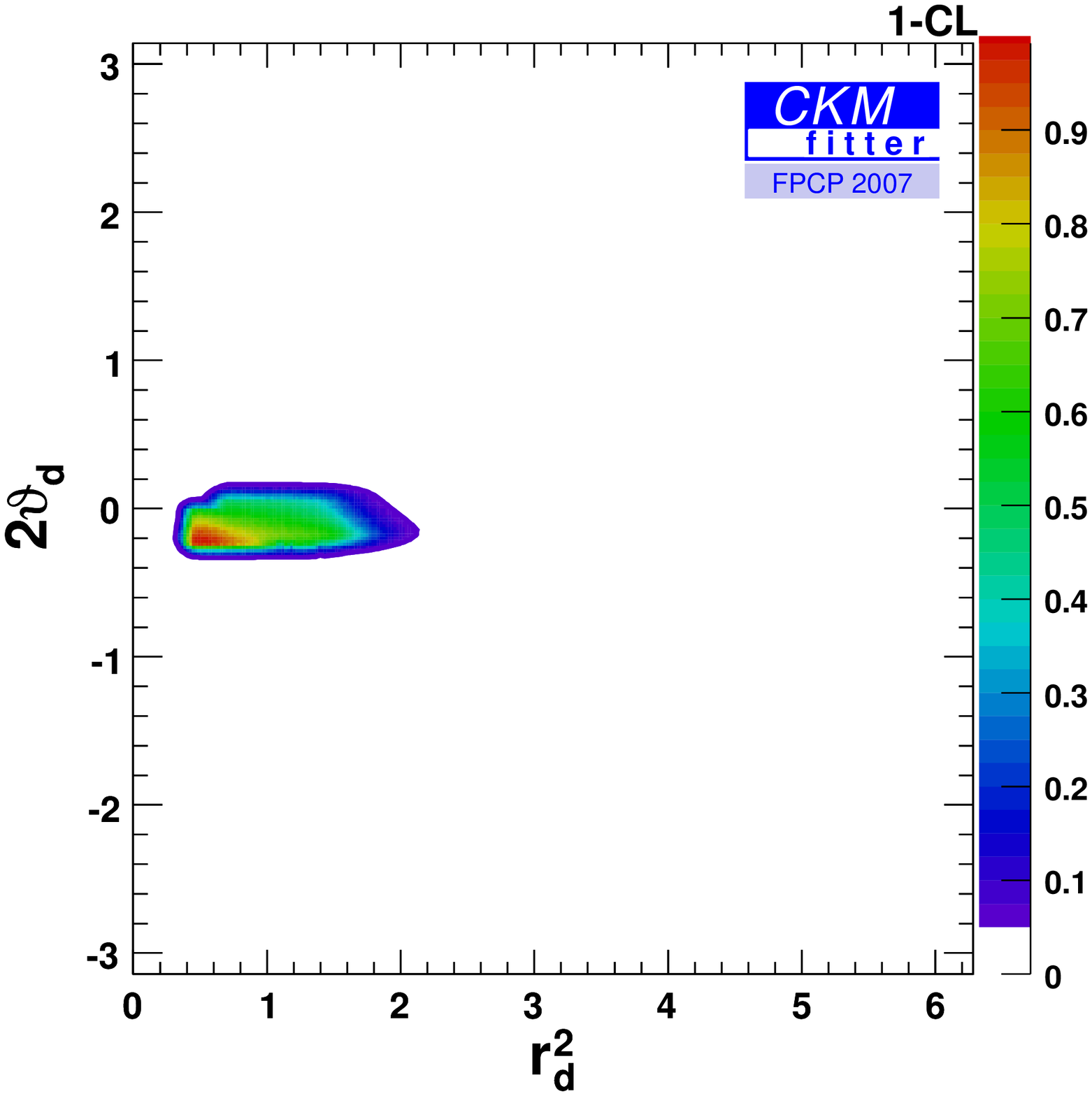}
\hspace{2em}
\includegraphics[width=2.85in]{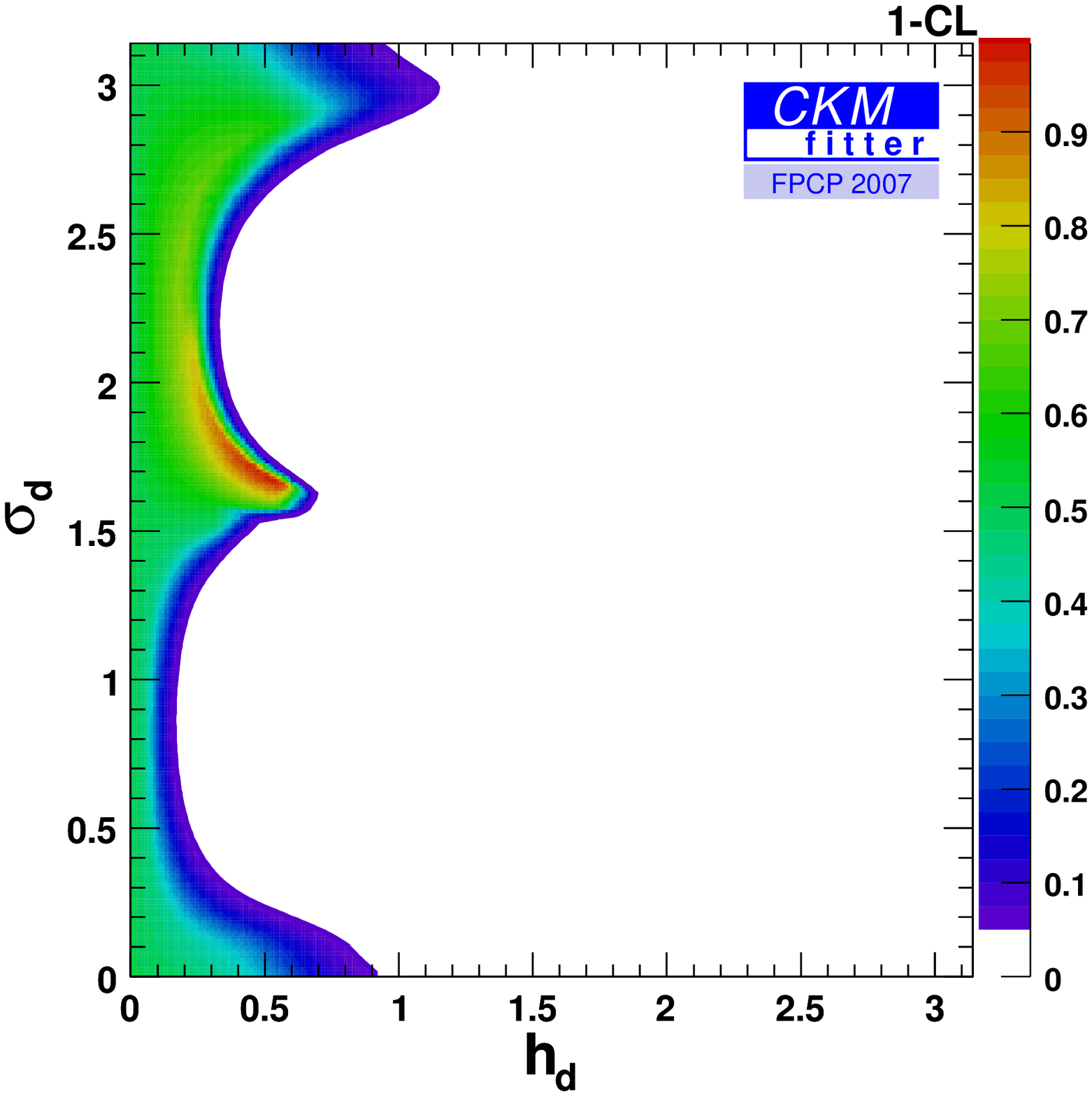}
\caption{Constraints in the (a) $r_d^2-2\theta_d$ plane, and (b)
  $h_d-\sigma_d$ plane, assuming that NP contributions to tree
  level processes are negligible \cite{ckmfitter}.}
\label{fig:rdtd}
\end{center}
\end{figure}

All that we need to do in order to establish whether the SM dominates
the observed CP violation, and to put an upper bound on the new
physics contribution to $B^0-\overline{B}{}^0$ mixing, is to project
the results of the fit performed in the previous subsection on the
$r_d^2-2\theta_d$ plane. If we find that $\theta_d\ll\beta$, then the
SM dominance in the observed CP violation will be established.
The constraints are shown in Fig.~\ref{fig:rdtd}(a). Indeed,
$\theta_d\ll\beta$.

An alternative way to present the data is to use the $h_d,\sigma_d$
parametrization,
\beq
r_d^2e^{2i\theta_d}=1+h_d e^{2i\sigma_d}.
\eeq
While the $r_d,\theta_d$ parameters give the relation between the full
mixing amplitude and the SM one, and are convenient to apply to the
measurements, the $h_d,\sigma_d$ parameters give the relation between
the new physics and SM contributions, and are more convenient in
testing theoretical models:
\beq\label{defhsig}
h_de^{2i\sigma_d}=\frac{M_{12}^{\rm NP}}{M_{12}^{\rm SM}}.
\eeq
The constraints in the $h_d-\sigma_d$ plane are shown in
Fig.~\ref{fig:rdtd}(b). We can make the following two statements:
\begin{enumerate}
  \item A new physics contribution to $B^0-\overline{B}^0$ mixing
      amplitude that carries a phase that is significantly different
      from the KM phase is constrained to lie below the 20-30\% level.
 \item A new physics contribution to the $B^0-\overline{B}^0$ mixing
   amplitude which is aligned with the KM phase is constrained to be
   at most comparable to the CKM contribution.
 \end{enumerate}
 One can reformulate these statements as follows:
 \begin{enumerate}
   \item The KM mechanism dominates CP violation in
     $B^0-\overline{B}^0$ mixing.
     \item The CKM mechanism is a major player in $B^0-\overline{B}^0$
       mixing.
     \end{enumerate}

\section{The new physics flavor puzzle}
\label{sec:nppuzzle}
%
\subsection{A model independent discussion}
It is clear that the Standard Model is not a complete theory of
Nature:
\begin{enumerate}
\item It does not include gravity, and therefore it cannot be valid at
  energy scales above $m_{\rm Planck}\sim10^{19}$ GeV:
\item It does not allow for neutrino masses, and therefore it cannot
  be valid at energy scales above $m_{\rm seesaw}\sim10^{15}$ GeV;
\item The fine-tuning problem of the Higgs mass suggests that the scale where the SM is replaced with a more fundamental theory is actually much lower, $m_{\rm top-partners}\lsim$ a few TeV.
\item If the dark matter is made of weakly interacting massive particles (WIMPs) then, again, a low scale of new physics is likely, $m_{\rm wimp}\lsim$ a few TeV.
\end{enumerate}
Given that the SM is only an effective low energy theory,
non-renormalizable terms must be added to ${\cal L}_{\rm SM}$ of Eq.
(\ref{LagSM}). These are terms of dimension higher than four in the
fields which, therefore, have couplings that are inversely
proportional to the scale of new physics $\Lambda_{\rm NP}$. For
example, the lowest dimension non-renormalizable terms are dimension
five:
\beq\label{Hnint}
-{\cal L}_{\rm Yukawa}^{\rm dim-5}=
{Z_{ij}^\nu\over \Lambda_{\rm
    NP}}L_{Li}^I L_{Lj}^I\phi\phi+{\rm h.c.}.
\eeq
These are the seesaw terms, leading to neutrino masses.

{\bf Exercise 5:} {\it How does the global symmetry breaking pattern
(\ref{globre}) change when (\ref{Hnint}) is taken into account?}

{\bf Exercise 6:} {\it What is the number of physical lepton flavor
parameters in this case? Identify these parameters in the mass basis.}

As concerns quark flavor physics, consider, for example, the following
dimension-six, four-fermion, flavor changing operators:
\beq\label{eq:ffll}
{\cal L}_{\Delta F=2}=
\frac{z_{sd}}{\Lambda_{\rm NP}^2}(\overline{d_L}\gamma_\mu s_L)^2
+\frac{z_{cu}}{\Lambda_{\rm NP}^2}(\overline{c_L}\gamma_\mu u_L)^2
+\frac{z_{bd}}{\Lambda_{\rm NP}^2}(\overline{d_L}\gamma_\mu b_L)^2
+\frac{z_{bs}}{\Lambda_{\rm NP}^2}(\overline{s_L}\gamma_\mu b_L)^2.
\eeq
Each of these terms contributes to the mass splitting between the
corresponding two neutral mesons. For example, the term ${\cal
  L}_{\Delta B=2}\propto(\overline{d_L}\gamma_\mu b_L)^2$ contributes
to $\Delta m_B$, the mass difference between the two neutral
$B$-mesons. We use $M_{12}^B=\frac{1}{2m_B}\langle B^0|{\cal
  L}_{\Delta F=2}|\overline{B}^0\rangle$ and
\beq
\langle B^0|(\overline{d_{La}}\gamma^\mu
b_{La})(\overline{d_{Lb}}\gamma_\mu b_{Lb})|\overline{B}^0\rangle =
-\frac13 m_B^2f_B^2 B_B.
\eeq
This leads to $\Delta m_B/m_B=2|M_{12}^B|/m_B\sim (|z_{bd}|/3)(f_B/\Lambda_{\rm NP})^2$.
Analogous expressions hold for the other neutral mesons.

The experimental results for CP conserving and CP violating observables related to  neutral meson mixing (mass splittings and CP asymmetries in tree level decays,
respectively) are given in Table \ref{tab:monetwoexp}.

\begin{table}[t]
  \caption{Measurements related to neutral meson mixing}
\label{tab:monetwoexp}
\begin{center}
\begin{tabular}{ccc} \hline\hline
  \rule{0pt}{1.2em}%
  Sector & CP-conserving & CP-violating
  \cr \hline
sd &\ $\Delta m_K/m_K=7.0\times10^{-15}$\ &\ $\epsilon_K=2.3\times10^{-3}$  \cr
cu &\ $\Delta m_D/m_D=8.7\times10^{-15}$\ &\ $A_\Gamma/y_{\rm CP}\lsim0.2$  \cr
bd &\ $\Delta m_B/m_B=6.3\times10^{-14}$\ &\ $S_{\psi K}=+0.67\pm0.02$  \cr
bs &\ $\Delta m_{B_s}/m_{B_s}=2.1\times10^{-12}$\ &\ $S_{\psi\phi}=-0.04\pm0.09$  \cr
\hline\hline
\end{tabular}
\end{center}
\end{table}

The measurements quoted in Table \ref{tab:monetwoexp} lead, for a given value of $|z_{ij}|$ and $z^I_{ij}\equiv {\cal I}m(z_{ij})$, to lower bounds on the scale $\Lambda_{\rm NP}$. In Table \ref{tab:lnpbounds} we give the bounds that correspond to $|z_{ij}|=1$ and to $z^I_{ij}=1$. The bounds scale like $\sqrt{z_{ij}}$ and  $\sqrt{z^I_{ij}}$, respectively.

\begin{table}[t]
  \caption{Lower bounds on the scale of new physics $\Lambda_{\rm NP}$, in units of TeV. The bounds from CP conserving (violating) observables scale like $\sqrt{z_{ij}}$ ($\sqrt{z^I_{ij}}$).}
\label{tab:lnpbounds}
\begin{center}
\begin{tabular}{ccc} \hline\hline
  \rule{0pt}{1.2em}%
  $ij$ & CP-conserving & CP-violating
  \cr \hline
sd &\ $1\times10^3$\ &\ $2\times10^4$  \cr
cu &\ $1\times10^3$\ &\ $3\times10^3$  \cr
bd &\ $4\times10^2$\ &\ $8\times10^2$  \cr
bs &\ $7\times10^1$\ &\ $2\times10^2$  \cr
\hline\hline
\end{tabular}
\end{center}
\end{table}

We conclude that if the new physics has a generic flavor structure, that is
$z_{ij}={\cal O}(1)$, then its scale must be above $10^3-10^4$ TeV.
If the leading contributions involve electroweak loops, the lower bound is somewhat lower, of order $10^2-10^3$ TeV. The bounds from the corresponding four-fermi
terms with LR structure, instead of the LL structure of Eq. (\ref{eq:ffll}),
are even stronger. {\it If indeed $\Lambda_{\rm NP}\gg TeV$, it means
that we have misinterpreted the hints from the fine-tuning problem
and the dark matter puzzle.}

There is, however, another way to look at these constraints:
\beqa\label{zconsa}
z_{sd}&\lsim&8\times10^{-7}\ (\Lambda_{\rm NP}/TeV)^2,\no\\
z_{cu}&\lsim&5\times10^{-7}\ (\Lambda_{\rm NP}/TeV)^2,\no\\
z_{bd}&\lsim&5\times10^{-6}\ (\Lambda_{\rm NP}/TeV)^2,\no\\
z_{bs}&\lsim&2\times10^{-4}\ (\Lambda_{\rm NP}/TeV)^2,
\eeqa
\beqa\label{zconsi}
z_{sd}^I&\lsim&6\times10^{-9}\ (\Lambda_{\rm NP}/TeV)^2,\no\\
z_{cu}^I&\lsim&1\times10^{-7}\ (\Lambda_{\rm NP}/TeV)^2,\no\\
z_{bd}^I&\lsim&1\times10^{-6}\ (\Lambda_{\rm NP}/TeV)^2,\no\\
z_{bs}^I&\lsim&2\times10^{-5}\ (\Lambda_{\rm NP}/TeV)^2.
\eeqa
{\it It could be that the scale of new physics is of order TeV, but
  its flavor structure is far from generic.} Specifically,
if new particles at the TeV scale couple to the SM fermions, then there
are two ways in which their contributions to FCNC processes, such as neutral meson mixing, can be suppressed: degeneracy and alignment. Either of these principles, or a combination of both, signifies non-generic structure.

One can use the language of effective operators also for the SM,
integrating out all particles significantly heavier than the neutral
mesons (that is, the top, the Higgs and the weak gauge bosons). Thus,
the scale is $\Lambda_{\rm SM}\sim m_W$. Since the leading
contributions to neutral meson mixings come from box diagrams, the
$z_{ij}$ coefficients are suppressed by $\alpha_2^2$. To identify
the relevant flavor suppression factor, one can employ the spurion
formalism. For example, the flavor transition that is relevant to
$B^0-\overline{B}{}^0$ mixing involves $\overline{d_L}b_L$ which
transforms as $(8,1,1)_{SU(3)_q^3}$. The leading contribution must then
be proportional to $(Y^u Y^{u\dagger})_{13}\propto y_t^2
V_{tb}V_{td}^*$. Indeed, an explicit calculation, using VIA for the
matrix element and neglecting QCD corrections, gives (a detailed
derivation can be found in Appendix B of \cite{Branco:1999fs})
\beq \frac{2M_{12}^B}{m_B}\approx-\frac{\alpha_2^2}{12}
\frac{f_B^2}{m_W^2}S_0(x_t)(V_{tb}V_{td}^*)^2,
\eeq
where $x_i=m_i^2/m_W^2$ and
\beq
S_0(x)=\frac{x}{(1-x)^2}\left[1-\frac{11x}{4}+\frac{x^2}{4}-\frac{3x^2\ln
x}{2(1-x)}\right].  \eeq
Similar spurion analyses, or explicit calculations, allow us to
extract the weak and flavor suppression factors that apply in the
SM:
\beqa
{\cal I}m(z_{sd}^{\rm SM})&\sim&\alpha_2^2 y_t^2
|V_{td}V_{ts}|^2\sim1\times10^{-10},\no\\
z_{sd}^{\rm SM}&\sim&\alpha_2^2 y_c^2
|V_{cd}V_{cs}|^2\sim5\times10^{-9},\no\\
{\cal I}m(z_{cu}^{\rm SM})&\sim&\alpha_2^2 y_b^2
|V_{ub}V_{cb}|^2\sim2\times10^{-14},\no\\
z_{bd}^{\rm SM}&\sim&\alpha_2^2 y_t^2
|V_{td}V_{tb}|^2\sim7\times10^{-8},\no\\
z_{bs}^{\rm SM}&\sim&\alpha_2^2 y_t^2
|V_{ts}V_{tb}|^2\sim2\times10^{-6}.
\eeqa

Note that we did not include $z_{cu}^{\rm SM}$ in the list. The reason is tha it requires
a more detailed consideration. The naively leading short distance
contribution is $\propto \alpha_2^2(y_s^4/y_c^2)
|V_{cs}V_{us}|^2\sim5\times10^{-13}$. However, higher dimension terms
can replace a $y_s^2$ factor with $(\Lambda/m_D)^2$
\cite{Bigi:2000wn}. Moreover, long distance contributions are expected
to dominate. In particular, peculiar phase space effects
\cite{Falk:2001hx,Falk:2004wg} have been identified which are expected
to enhance $\Delta m_D$ to within an order of magnitude of the its
measured value. The CP violating part, on the other hand, is dominated by
short distance physics.

It is clear then that contributions from new physics at $\Lambda_{\rm
  NP}\sim1\ TeV$ should be suppressed by factors that are comparable
or smaller than the SM ones. Why does that happen? This is the new
physics flavor puzzle.

The fact that the flavor structure of new physics at the TeV scale
must be non-generic means that flavor measurements are a good probe of
the new physics. Perhaps the best-studied example is that of
supersymmetry. Here, the spectrum of the superpartners and the
structure of their couplings to the SM fermions will allow us to probe
the mechanism of dynamical supersymmetry breaking.

\subsection{The supersymmetric flavor puzzle}
\label{sec:dmix}
We consider, as an example, the contributions from the box diagrams
involving the squark doublets of the second and third generations,
$\tilde Q_{L2,3}$, to the $B_s-\overline{B_s}$ mixing amplitude. The
contributions are proportional to $K_{3i}^{d*} K^{d}_{2i}K_{3j}^{d*}
K^{d}_{2j}$, where $K^d$ is the mixing matrix of the gluino couplings
to a left-handed down quark and their supersymmetric squark partners
($\propto[(\delta^d_{LL})_{23}]^2$ in the mass insertion
approximation, described in Appendix \ref{app:susydel}). We work in
the mass basis for both quarks and squarks.  A detailed derivation
\cite{Raz:2002zx} is given in Appendix \ref{app:susyd}. It gives:
\beqa\label{motsusyc}
M_{12}^s&=&\frac{\alpha_s^2m_{B_s}f_{B_s}^2B_{B_s}\eta_{\rm
    QCD}}{108m_{\tilde d}^2}
[11\tilde f_6(x)+4xf_6(x)]\frac{(\Delta\tilde m^2_{\tilde
    d})^2}{\tilde m_d^4} (K_{32}^{d*}K_{22}^{d})^2.
\eeqa
Here $m_{\tilde d}$ is the average mass of the two squark generations,
$\Delta m^2_{\tilde d}$ is the mass-squared difference, and
$x=m_{\tilde g}^2/m_{\tilde d}^2$.

Eq. (\ref{motsusyc}) can be translated into our generic language:
\beqa\label{eq:susyzbs}
\Lambda_{\rm NP}&=&m_{\tilde q},\\
z_1^{bs}&=&\frac{11\tilde f_6(x)+4x f_6(x)}{18}\alpha_s^2
\left(\frac{\Delta\tilde m_{\tilde d}^2}{m_{\tilde
      d}^2}\right)^2(K_{32}^{d*}K_{22}^{d})^2
\approx10^{-4}(\delta^{LL}_{23})^2,\no
\eeqa
where, for the last approximation, we took the example of $x=1$
[and used, correspondingly, $11\tilde f_6(1)+4f_6(1)=1/6$], and defined
\beq
\delta^{LL}_{23}=\left(\frac{\Delta\tilde m_{\tilde d}^2}{m_{\tilde
      d}^2}\right)(K_{32}^{d*}K_{22}^{d}).
\eeq
Similar expressions can be derived for the dependence of
$K^0-\overline{K^0}$ on $(\delta^d_{MN})_{12}$, $B^0-\overline{B^0}$
on $(\delta^d_{MN})_{13}$, and $D^0-\overline{D^0}$ on
$(\delta^u_{MN})_{12}$. Then we can use the constraints of Eqs.
(\ref{zconsa},\ref{zconsi}) to put upper bounds on
$(\delta^q_{MN})_{ij}$. Some examples are given in Table
\ref{tab:susydel} (see Ref. \cite{Isidori:2010kg} for details and list
of references).

\begin{table}[t]
  \caption{The phenomenological upper bounds on $(\delta^q_{LL})_{ij}$
    and
    $\langle\delta^q_{ij}\rangle=\sqrt{(\delta^q_{LL})_{ij}(\delta^q_{RR})_{ij}}$.
    Here $q=u,d$ and $M=L,R$. The constraints are given for $m_{\tilde
    q}=1$ TeV and $x=m_{\tilde g}^2/m_{\tilde q}^2=1$. We assume that
    the phases could suppress the imaginary part by a factor of
    $\sim0.3$. Taken from Ref. \cite{Isidori:2010kg}.}
\label{tab:susydel}
\begin{center}
\begin{tabular}{cc|cc} \hline\hline
  \rule{0pt}{1.2em}%
  $q$ & $ij$ & $(\delta^q_{LL})_{ij}$ & $\langle\delta^q_{ij}\rangle$
  \cr \hline
d & 12 & 0.03 & 0.002  \cr
d & 13 & 0.2 & 0.07  \cr
d & 23 & 0.2 & 0.07   \cr
u & 12 & 0.1 & 0.008  \cr
\hline\hline
\end{tabular}
\end{center}
\end{table}

We learn that, in most cases, we need $\delta^q_{ij}/m_{\tilde
  q}\ll1/{\rm TeV}$.  One can immediately identify three generic ways
in which supersymmetric contributions to neutral meson mixing can be
suppressed:
\begin{enumerate}
\item Heaviness: $m_{\tilde q}\gg1\ TeV$;
\item Degeneracy: $\Delta m^2_{\tilde q}\ll m_{\tilde q}^2$;
\item Alignment: $K^{q}_{ij}\ll1$.
\end{enumerate}
When heaviness is the only suppression mechanism, as in split
supersymmetry \cite{ArkaniHamed:2004fb}, the squarks are very heavy
and supersymmetry no longer solves the fine tuning
problem. (When the first two squark generations are
mildly heavy and the third generation is light, as in effective
supersymmetry \cite{Cohen:1996vb}, the
fine tuning problem is still solved, but additional suppression
mechanisms are needed.) If we want to maintain supersymmetry as a
solution to the fine tuning problem, either degeneracy or alignment or
a combination of the two is needed. This means that the flavor structure
of supersymmetry is not generic, as argued in the previous section.

Take, for example, $(\delta^d_{LL})_{12}\leq 0.03$. Naively, one might
expect the alignment to be of order $(V_{cd}V_{cs}^*)\sim0.2$, which
is far from sufficient by itself. Barring a very precise alignment ($|K^d_{12}|\ll|V_{us}|$) \cite{Nir:1993mx,Leurer:1993gy}
and accidental cancelations, we are led to conclude that the first
two squark generations must be quasi-degenerate. Actually, by
combining the constraints from $K^0-\overline{K^0}$ mixing and
$D^0-\overline{D^0}$ mixing,  one can show that this is the case
independently of assumptions about the alignment
\cite{Ciuchini:2007cw,Nir:2007ac,Gedalia:2012pi}. Analogous conclusions can be drawn
for many TeV-scale new physics scenarios: a strong level of degeneracy
is required (for definitions and detailed analysis, see
\cite{Blum:2009sk}).

{\bf Exercise 9:} {\it Does $K_{31}^d\sim|V_{ub}|$ suffice to satisfy the
$\Delta m_B$ constraint with neither degeneracy nor heaviness? (Use
the two generation approximation and ignore the second generation.)}

Is there a natural way to make the squarks degenerate?  Degeneracy
requires that the $3\times3$ matrix of soft supersymmetry breaking
mass-squared terms $\tilde m^2_{Q_L}\simeq\tilde m^2_{\tilde q}{\bf
  1}$. We have mentioned already that flavor universality is a generic
feature of gauge interactions. Thus, the requirement of degeneracy is
perhaps a hint that supersymmetry breaking is {\it gauge mediated} to
the MSSM fields.

\subsection{Minimal flavor violation (MFV)}
\label{sec:mfv}
If supersymmetry breaking is gauge mediated, the squark mass matrices
for $SU(2)_L$- doublet and $SU(2)_L$-singlet squarks have
the following form at the scale of mediation $m_M$:
\beqa\label{mllgm}
\tilde M^2_{U_L}(m_M)&=&\left(m^2_{\tilde Q_L}+D_{U_L}\right){\bf
  1}+M_u M_u^\dagger,\no\\
\tilde M^2_{D_L}(m_M)&=&\left(m^2_{\tilde Q_L}+D_{D_L}\right){\bf
  1}+M_d M_d^\dagger,\no\\
\tilde M^2_{U_R}(m_M)&=&\left(m^2_{\tilde U_R}+D_{U_R}\right){\bf
  1}+M_u^\dagger M_u,\no\\
\tilde M^2_{D_R}(m_M)&=&\left(m^2_{\tilde D_R}+D_{D_R}\right){\bf
  1}+M_d^\dagger M_d,
\eeqa
where $D_{q_A}=(T_3)_{q_A}-(Q_{\rm EM})_{q_A}s^2_W m_Z^2\cos2\beta$
are the $D$-term contributions. Here, the only source of the
$SU(3)^3_q$ breaking are the SM Yukawa matrices.

This statement holds also when the renormalization group evolution
is applied to find the form of these matrices at the weak scale.
Taking the scale of the soft breaking terms $m_{\tilde q_A}$ to be
somewhat higher than the
electroweak breaking scale $m_Z$ allows us to neglect the
$D_{q_A}$ and $M_q$ terms in (\ref{mllgm}). Then we obtain
\beqa\label{mllrrmz}
\tilde M^2_{Q_L}(m_Z)&\sim&m^2_{\tilde Q_L}\left(r_3{\bf 1}+c_u
  Y_uY_u^\dagger+c_d Y_d Y_d^\dagger\right),\no\\
\tilde M^2_{U_R}(m_Z)&\sim&m^2_{\tilde U_R}\left(r_3{\bf 1}+c_{uR}
  Y_u^\dagger Y_u\right),\no\\
\tilde M^2_{D_R}(m_Z)&\sim&m^2_{\tilde D_R}\left(r_3{\bf 1}+c_{dR}
  Y_d^\dagger Y_d\right).
\eeqa
Here $r_3$ represents the universal RGE contribution that is proportional
to the gluino mass ($r_3={\cal O}(6)\times(M_3(m_M)/m_{\tilde q}(m_M))$)
and the $c$-coefficients depend logarithmically on $m_M/m_Z$ and can be
of ${\cal O}(1)$ when $m_M$ is not far below the GUT scale.

Models of gauge mediated supersymmetry breaking (GMSB) provide a
concrete example of a large class of models that obey a simple
principle called {\it minimal flavor violation} (MFV)
\cite{D'Ambrosio:2002ex}. This principle guarantees that low energy
flavor changing processes deviate only very little from the SM
predictions.  The basic idea can be described as follows. The gauge
interactions of the SM are universal in flavor space. The only
breaking of this flavor universality comes from the three Yukawa
matrices, $Y^u$, $Y^d$ and $Y^e$. If this remains true in the presence
of the new physics, namely $Y^u$, $Y^d$ and $Y^e$ are the only flavor
non-universal parameters, then the model belongs to the MFV class.

Let us now formulate this principle in a more formal way, using the
language of spurions that we presented in section \ref{sec:spurions}.
The Standard Model with vanishing Yukawa couplings has a large global
symmetry (\ref{gglobal},\ref{susuu}). In this section we concentrate
only on the quarks. The non-Abelian part of the flavor symmetry for
the quarks is $SU(3)_q^3$ of Eq. (\ref{susuu}) with
the three generations of quark fields transforming as follows:
\beq
Q_L(3,1,1),\ \ U_R(1,3,1),\ \ D_R(1,1,3).
\eeq
The Yukawa interactions,
\beq\label{eq:lagy}
{\cal L}_Y=\overline{Q_L}Y^d D_R H + \overline{Q_L}Y^u U_R H_c ,
\eeq
($H_c=i\tau_2 H^*$) break this symmetry. The Yukawa couplings can thus
be thought of as spurions with the following transformation
properties under $SU(3)_q^3$ [see Eq. (\ref{Gglobq})]:
\beq
Y^u\sim(3,\bar3,1),\qquad Y^d\sim(3,1,\bar3).
\eeq
When we say ``spurions'', we mean that we pretend that the Yukawa
matrices are fields which transform under the flavor symmetry, and
then require that all the Lagrangian terms, constructed
from the SM fields, $Y^d$ and $Y^u$, must be (formally)
invariant under the flavor group $SU(3)_q^3$. Of course, in reality,
${\cal L}_Y$ breaks $SU(3)_q^3$ precisely because $Y^{d,u}$ are {\it
  not} fields and do not transform under the symmetry.

The idea of minimal flavor violation is relevant to extensions of the
SM, and can be applied in two ways:
\begin{enumerate}
 \item If we consider the SM as a low energy effective theory, then
   all higher-dimension operators, constructed from SM-fields and
   $Y$-spurions, are formally invariant under $G_{\rm global}$.
\item If we consider a full high-energy theory that extends the SM,
  then all operators, constructed from SM and the new fields, and from
  $Y$-spurions, are formally invariant under $G_{\rm global}$.
\end{enumerate}

{\bf Exercise 10:} {\it Use the spurion formalism to argue that, in MFV
models, the $K_L\to\pi^0\nu\bar\nu$ decay amplitude is proportional to
$y_t^2 V_{td}V_{ts}^*$.}

{\bf Exercise 11:} {\it Find the flavor suppression factors in the
  $z_i^{bs}$ coefficients, if MFV is imposed, and compare to the bounds
  in Eq. (\ref{zconsa}).}

Examples of MFV models include models of supersymmetry with
gauge-mediation or with anomaly-mediation of its breaking.

\subsubsection{Testing MFV at the LHC}
If the LHC discovers new particles that couple to the SM fermions,
then it will be able to test solutions to the new physics flavor
puzzle such as MFV \cite{Grossman:2007bd}. Much of its power to test
such frameworks is based on identifying top and bottom quarks.

To understand this statement, we notice that the spurions $Y^u$ and
$Y^d$ can always be written in terms of the two diagonal Yukawa
matrices $\lambda_u$ and $\lambda_d$ and the CKM matrix $V$, see
Eqs. (\ref{speint},\ref{deflamd}). Thus, the only source of quark
flavor changing transitions in MFV models is the CKM matrix. Next,
note that to an accuracy that is better than ${\cal O}(0.05)$, we can
write the CKM matrix as follows:
\beq\label{ckmapp}
V=\begin{pmatrix} 1&0.23&0\\ -0.23&1&0\\ 0&0&1\\ \end{pmatrix}.
\eeq

{\bf Exercise 12:} {\it The approximation (\ref{ckmapp}) should be
intuitively obvious to top-physicists, but definitely
counter-intuitive to bottom-physicists. (Some of them have dedicated a
large part of their careers to experimental or theoretical efforts to
determine $V_{cb}$ and $V_{ub}$.) What does the approximation
imply for the bottom quark? When we take into account that it is
only good to ${\cal O}(0.05)$, what would the implications be?}

We learn that the third generation of quarks is decoupled, to a good
approximation, from the first two. This, in turn, means that any new
particle that couples to an odd number of the SM quarks (think, for
example, of heavy quarks in vector-like representations of $G_{\rm
  SM}$), decay into either third generation quark, or to non-third
generation quark, but not to both. For example, in Ref.
\cite{Grossman:2007bd}, MFV models with additional charge $-1/3$,
$SU(2)_{\rm L}$-singlet quarks -- $B^\prime$ -- were considered. A
concrete test of MFV was proposed, based on the fact that the largest
mixing effect involving the third generation is of order
$|V_{cb}|^2\sim0.002$: Is the following prediction, concerning events
of $B^\prime$ pair production, fulfilled:
\beq
\frac{\Gamma(B^\prime\overline{B^\prime}\to Xq_{1,2}q_3)}
{\Gamma(B^\prime\overline{B^\prime}\to Xq_{1,2}q_{1,2})+
  \Gamma(B^\prime\overline{B^\prime}\to Xq_3q_3)}\lsim10^{-3}.
\eeq
If not, then MFV is excluded. One could similarly test various versions of minimal lepton flavor violation (MLFV) \cite{Cirigliano:2005ck,Cirigliano:2006su,Cirigliano:2006nu,Branco:2006hz,Chen:2008qg,Gross:2010ce}.

Analogous tests can be carried out in the supersymmetric framework
\cite{Feng:2007ke,Hiller:2008wp,Hiller:2008sv,Feng:2009bs,Feng:2009yq,Feng:2009bd,Hiller:2010dv}.
Here, there is also a generic prediction that, in each of the three
sectors ($Q_L,U_R,D_R$), squarks of the first two generations are
quasi-degenerate, and do not decay into third generation quarks.
Squarks of the third generation can be separated in mass (though, for
small $\tan\beta$, the degeneracy in the $\tilde D_R$ sector is
threefold), and decay only to third generation quarks.

We conclude that measurements at the LHC related to new particles that
couple to the SM fermions are likely to teach us much more about flavor
physics.

\section{The Standard Model flavor puzzle}
The SM has thirteen flavor parameters: six quark Yukawa couplings, four CKM parameters (three angles and a phase), and three charged lepton Yukawa couplings. (One can use fermions masses instead of the fermion Yukawa couplings, $Y_f=\sqrt{2}m_f/v$.) The orders of magnitudes of these thirteen dimensionless parameters are as follows:
\beqa\label{eq:yukoom}
Y_t&\sim&1,\ \ \ Y_c\sim10^{-2},\ \ \ Y_u\sim10^{-5},\nonumber\\
Y_b&\sim&10^{-2},\ \ \ Y_s\sim10^{-3},\ \ \ Y_d\sim10^{-4},\nonumber\\
Y_\tau&\sim&10^{-2},\ \ \ Y_\mu\sim10^{-3},\ \ \ Y_e\sim10^{-6},\nonumber\\
|V_{us}|&\sim&0.2,\ \ |V_{cb}|\sim0.04,\ \ |V_{ub}|\sim0.004,\ \ \ \delta_{\rm KM}\sim1.
\eeqa
Only two of these parameters are clearly of ${\cal O}(1)$, the top-Yukawa and the KM phase. The other flavor parameters exhibit smallness and hierarchy. Their values span six orders of magnitude. It may be that this set of numerical values are just accidental. More likely, the smallness and the hierarchy have a reason. The question of why there is smallness and hierarchy in the SM flavor parameters constitutes ``The Standard Model flavor puzzle."

The motivation to think that there is indeed a structure in the flavor parameters is strengthened by considering the values of the four SM parameters that are not flavor parameters, namely the three gauge couplings and the Higgs self-coupling:
\beq
g_s\sim1,\ \ g\sim0.6,\ \ e\sim0.3,\ \ \lambda\sim0.2.
\eeq
This set of values does seem to be a random distribution of order-one numbers, as one would naively expect.

A few examples of mechanisms that were proposed to explain the observed structure of the flavor parameters are the following:
\begin{itemize}
\item An approximate Abelian symmetry (``The Froggatt-Nielsen mechanism" \cite{Froggatt:1978nt});
\item An approximate non-Abelian symmetry (see {\it e.g.} \cite{Dine:1993np});
\item Conformal dynamics (``The Nelson-Strassler mechanism" \cite{Nelson:2000sn});
\item Location in an extra dimension \cite{ArkaniHamed:1999dc}.
\end{itemize}
We will take as an example the Froggatt-Nielsen mechanism.

\subsection{The Froggatt-Nielsen mechanism}
\label{sec:fn}
Small numbers and hierarchies are often explained by approximate symmetries. For example, the small mass splitting between the charged and neural pions finds an explanation in the approximate isospin (global $SU(2)$) symmetry of the strong interactions.

Approximate symmetries lead to selection rules which account for the size of deviations from the symmetry limit. Spurion analysis is particularly convenient to derive such selection rules. The Froggatt-Nielsen mechanism postulates a $U(1)_H$ symmetry, that is broken by a small spurion $\epsilon_H$. Without loss of generality, we assign $\epsilon_H$ a $U(1)_H$ charge of $H(\epsilon_H)=-1$. Each SM field is assigned a $U(1)_H$ charge. In general, different fermion generations are assigned different charges, hence the term `horizontal symmetry.' The rule is that each term in the Lagrangian, made of SM fields and the spurion should be formally invariant under $U(1)_H$.

The approximate $U(1)_H$ symmetry thus leads to the following selection rules:
\beqa\label{eq:selrul}
Y^u_{ij}&=&\epsilon_H^{|H(\bar Q_i)+H(U_j)+H(\phi_u)|},\no\\
Y^d_{ij}&=&\epsilon_H^{|H(\bar Q_i)+H(D_j)+H(\phi_d)|},\no\\
Y^e_{ij}&=&\epsilon_H^{|H(\bar L_i)+H(E_j)-H(\phi_d)|}.
\eeqa

As a concrete example, we take the following set of charges:
\beqa
H(\bar Q_i)&=&H(U_i)=H(E_i)=(2,1,0),\no\\
H(\bar L_i)&=&H(D_i)=(0,0,0),\no\\
H(\phi_u)&=&H(\phi_d)=0.
\eeqa
It leads to the following parametric suppressions of the Yukawa couplings:
\beq
Y^u\sim \begin{pmatrix}\epsilon^4 & \epsilon^3 & \epsilon^2 \\
\epsilon^3 & \epsilon^2 & \epsilon \\ \epsilon^2 & \epsilon & 1\\ \end{pmatrix},\ \
Y^d\sim (Y^e)^T\sim \begin{pmatrix}\epsilon^2 & \epsilon^2 & \epsilon^2 \\
\epsilon & \epsilon & \epsilon \\ 1 & 1 & 1\\ \end{pmatrix}.
\eeq
We emphasize that for each entry we give the parametric suppression (that is the power of $\epsilon$), but each entry has an unknown (complex) coefficient of order one, and there are no relations between the order one coefficients of different entries.

The structure of the Yukawa matrices dictates the parametric suppression of the physical observables:
\beqa\label{eq:yukfn}
Y_t&\sim&1,\ \ \ Y_c\sim\epsilon^2,\ \ \ Y_u\sim\epsilon^4,\nonumber\\
Y_b&\sim&1,\ \ \ Y_s\sim\epsilon,\ \ \ Y_d\sim\epsilon^2,\nonumber\\
Y_\tau&\sim&1,\ \ \ Y_\mu\sim\epsilon,\ \ \ Y_e\sim\epsilon^2,\nonumber\\
|V_{us}|&\sim&\epsilon,\ \ |V_{cb}|\sim\epsilon,\ \ |V_{ub}|\sim\epsilon^2,\ \ \ \delta_{\rm KM}\sim1.
\eeqa
For $\epsilon\sim0.05$, the parametric suppressions are roughly consistent with the observed hierarchy. In particular, this set of charges predicts that the down and charged lepton mass hierarchies are similar, while the up hierarchy is the square of the down hierarchy. These features are roughly realized in Nature.

{\bf Exercise 13:} {\it Derive the parametric suppression and approximate numerical values of $Y^u$, its eigenvalues, and the three angles of $V_L^u$, for $H(Q_i)=4,2,0$, $H(U_i)=3,2,0$ and $\epsilon_H=0.2$}

Could we explain any set of observed values with such an approximate symmetry? If we could, then the FN mechanism cannot be really tested. The answer however is negative. Consider, for example, the quark sector. Naively, we have 11 $U(1)_H$ charges that we are free to choose. However, the $U(1)_Y\times U(1)_B\times U(1)_{\rm PQ}$ symmetry implies that there are only 8 independent choices that affect the structure of the Yukawa couplings. On the other hand, there are 9 physical parameters. Thus, there should be a single relation between the physical parameters that is independent of the choice of charges. Assuming that the sum of charges in the exponents of Eq. (\ref{eq:selrul}) is of the same sign for all 18 combinations, the relation is
\beq
|V_{ub}|\sim|V_{us}V_{cb}|,
\eeq
which is fulfilled to within a factor of 2. There are also interesting inequalities (here $i<j$):
\beq
|V_{ij}|\gsim m(U_i)/m(U_j),\ m(D_i)/m(D_j).
\eeq
All six inequalities are fulfilled. Finally, if we order the up and the down masses from light to heavy, then the CKM matrix is predicted to be $\sim{\bf 1}$, namely the diagonal entries are not parametrically suppressed. This structure is also consistent with the observed CKM structure.

\subsection{The flavor of neutrinos}
Five neutrino flavor parameters have been measured in recent years (see {\it e.g.} \cite{GonzalezGarcia:2012sz}): two mass-squared differences,
\beq\label{eq:numass}
\Delta m^2_{21}=(7.5\pm0.2)\times10^{-5}\ {\rm eV}^2,\ \ \
|\Delta m^2_{32}|=(2.5\pm0.1)\times10^{-3}\ {\rm eV}^2,
\eeq
and the three mixing angles,
\beq\label{eq:numix}
|U_{e2}|=0.55\pm0.01,\ \ |U_{\mu3}|=0.64\pm0.02,\ \ |U_{e3}|=0.15\pm0.01.
\eeq
These parameters constitute a significant addition to the thirteen SM flavor parameters and provide, in principle, tests of various ideas to explain the SM flavor puzzle.

The numerical values of the parameters show various surprising features:
\begin{itemize}
\item $|U_{\mu3}|>{\rm any}\ |V_{ij}|$;
\item $|U_{e2}|>{\rm any}\ |V_{ij}|$;
\item $|U_{e3}|$ is not particularly small ($|U_{e3}|\not\ll|U_{e2}U_{\mu3}|$);
\item $m_2/m_3\gsim1/6>{\rm any}\ m_i/m_j$ for charged fermions.
\end{itemize}
These features can be summarized by the statement that, in contrast to the charged fermions, neither smallness nor hierarchy have been observed so far in the neutrino related parameters.

One way of interpretation of the neutrino data comes under the name of neutrino mass anarchy \cite{Hall:1999sn,Haba:2000be,deGouvea:2003xe}. It postulates  that the neutrino mass matrix has no structure, namely all entries are of the same order of magnitude. Normalized to an effective neutrino mass scale, $v^2/\Lambda_{\rm seesaw}$, the various entries are random numbers of order one. Note that anarchy means neither hierarchy nor degeneracy.

If true, the contrast between neutrino mass anarchy and quark and charged lepton mass hierarchy may be a deep hint for a difference between the flavor physics of Majorana and Dirac fermions. The source of both anarchy and hierarchy might, however, be explained by a much more mundane mechanism. In particular, neutrino mass anarchy could be a result of a FN mechanism, where the three left-handed lepton doublets carry the same FN charge. In that case, the FN mechanism predict parametric suppression of neither neutrino mass ratios nor leptonic mixing angles, which is quite consistent with (\ref{eq:numass}) and (\ref{eq:numix}). Indeed, the viable FN model presented in Section \ref{sec:fn} belongs to this class.

Another possible interpretation of the neutrino data is to take $m_2/m_3\sim|U_{e3}|\sim0.15$ to be small, and require that they are parametrically suppressed (while the other two mixing angles are order one). Such a situation is impossible to accommodate in a large class of FN models \cite{Amitai:2012pk}.

The same data, and in particular the proximity of $|U_{e2}|$ to $1/\sqrt{3}\simeq0.58$ and the proximity of $|U_{\mu3}|$ to $1/\sqrt2\simeq0.71$ led to a very different interpretation. This interpretation, termed `tribimaximal mixing' (TBM), postulates that the leptonic mixing matrix is parametrically close to the following special form \cite{Harrison:2002er}:
\beq\label{eq:tbm}
|U|_{\rm TBM}=\begin{pmatrix}
\frac{2}{\sqrt6} & \frac{1}{\sqrt3} & 0 \\
\frac{1}{\sqrt6} & \frac{1}{\sqrt3} & \frac{1}{\sqrt2} \\
\frac{1}{\sqrt6} & \frac{1}{\sqrt3} & \frac{1}{\sqrt2} \\ \end{pmatrix}.
\eeq
Such a form is suggestive of discrete non-Abelian symmetries, and indeed numerous models based on an $A_4$ symmetry have been proposed  \cite{Ma:2001dn,Altarelli:2010gt}. A significant feature of of TBM is that the third mixing angle should be close to $|U_{e3}|=0$. Until recently, there have been only upper bounds on $|U_{e3}|$, consistent with the models in the literature. In the last year, however, a value of $|U_{e3}|$ close to the previous upper bound has been established \cite{An:2012eh}, see Eq. (\ref{eq:numix}). Such a large value (and the consequent significant deviation of $|U_{\mu3}|$ from maximal bimixing) puts in serious doubt the TBM idea. Indeed, it is difficult in this framework, if not impossible, to account for $\Delta m^2_{12}/\Delta m^2_{23}\sim|U_{e3}|^2$ without fine-tuning \cite{Amitai:2012em}.

\section{Higgs physics: the new flavor arena}
A Higgs-like boson $h$  has been discovered by the ATLAS and CMS experiments at the LHC~\cite{:2012gk,:2012gu}. The fact that for the $f=\gamma\gamma$ and $f=ZZ^*$ final states, the experiments measure
\beq\label{eq:defmu}
R_f\equiv\frac{\sigma(pp\to h){\rm BR}(h\to f)}
{[\sigma(pp\to h){\rm BR}(h\to f)]^{\rm SM}},
\eeq
of order one (see {\it e.g.}~\cite{Carmi:2012zd}),
\beqa\label{eq:muexp}
R_{ZZ^*}&=&1.1\pm0.2,\\
R_{\gamma\gamma}&=&1.1\pm0.2,
\eeqa
is suggestive that the $h$-production via gluon-gluon fusion proceeds at a rate similar to the Standard Model (SM) prediction, giving a strong indication that $Y_t$, the $ht\bar t$ Yukawa coupling, is of order one. This first determination of $Y_t$ signifies a new arena for the exploration of {\it flavor physics}.

In the future, measurements of $R_{b\bar b}$ and $R_{\tau^+\tau^-}$ will allow us to extract additional flavor parameters: $Y_b$, the $hb\bar b$ Yukawa coupling, and $Y_\tau$, the $h\tau^+\tau^-$ Yukawa coupling. For the latter, the current allowed range is already quite restrictive:
\beq
R_{\tau^+\tau^-}=1.0\pm0.4.
\eeq
It may well be that the values of $Y_b$ and/or $Y_\tau$ will deviate from their SM values. The most likely explanation of such deviations will be that there are more than one Higgs doublets, and that the doublet(s) that couple to the down and charged lepton sectors are not the same as the one that couples to the up sector.

A more significant test of our understanding of flavor physics, which might provide a window into new flavor physics, will come further in the future,
when $R_{\mu^+\mu^-}$ is measured. (At present, there is an upper bound, $R_{\mu^+\mu^-}<9.8$.) The ratio
\beq\label{eq:rmutau}
X_{\mu^+\mu^-}\equiv\frac{{\rm BR}(h\to\mu^+\mu^-)}{{\rm BR}(h\to\tau^+\tau^-)},
\eeq
is predicted within the SM with impressive theoretical cleanliness. To leading order, it is given by $X_{\mu^+\mu^-}=m_\mu^2/m_\tau^2$, and the corrections of order $\alpha_W$ and of order $m_\mu^2/m_\tau^2$ to this leading result are known. It is an interesting question to understand what can be learned from a test of this relation \cite{Dery:2013rta,Dery:2013aba}.

It is also possible to search for the SM-forbidden decay modes, $h\to\mu^\pm\tau^\mp$ \cite{Blankenburg:2012ex,Harnik:2012pb,Davidson:2012ds,Arhrib:2012ax}. A measurement of, or an upper bound on
\beq\label{eq:brhmt}
X_{\mu\tau}\equiv\frac{{\rm BR}(h\to\mu^+\tau^-)+{\rm BR}(h\to\mu^-\tau^+)}{{\rm BR}(h\to\tau^+\tau^-)},
\eeq
would provide additional information relevant to flavor physics. Thus, a broader question is to understand the implications for flavor physics of measurements of $R_{\tau^+\tau^-}$, $X_{\mu^+\mu^-}$ and $X_{\mu\tau}$ \cite{Dery:2013rta}.

Let us take as an example how we can use the set of these three measurements if there is a single light Higgs boson. A violation of the SM relation $Y_{ij}^{\rm SM}=\frac{\sqrt{2} m_i}{v}\delta_{ij}$, is a consequence of nonrenormalizable terms. The leading ones are the $d=6$ terms. In the interaction basis, we have
\beqa\label{eq:dsix}
{\cal L}_Y^{d=4}&=&-\lambda_{ij}\bar f_L^i f_R^j \phi+{\rm h.c.},\\
{\cal L}_Y^{d=6}&=&-\frac{\lambda^\prime_{ij}}{\Lambda^2}\bar f_L^i f_R^j \phi
(\phi^\dagger \phi)+{\rm h.c.}\,,\nonumber
\eeqa
where expanding around the vacuum we have $\phi=(v+h)/\sqrt2$. Defining $V_{L,R}$ via
\beq\label{eq:vlr}
\sqrt2 m=V_L\left(\lambda+\frac{v^2}{2\Lambda^2}\lambda^\prime\right)
V_R^\dagger v,
\eeq
where $m={\rm diag}(m_e,m_\mu,m_\tau)$, and defining $\hat\lambda$ via
\beq\label{eq:deflamh}
\hat\lambda=V_L\lambda^\prime V_R^\dagger,
\eeq
we obtain
\beq\label{eq:yneqm}
Y_{ij}=\frac{\sqrt2 m_i}{v}\delta_{ij}+\frac{v^2}{\Lambda^2}\hat\lambda_{ij}.
\eeq

To proceed, one has to make assumptions about the structure of $\hat\lambda$. In what follows, we consider first the assumption of minimal flavor violation (MFV) and then a Froggatt-Nielsen (FN) symmetry.

\subsection{MFV}
MFV requires that the leptonic part of the Lagrangian is invariant under an $SU(3)_L\times SU(3)_E$ global symmetry, with the left-handed lepton doublets transforming as $(3,1)$,
the right-handed charged lepton singlets transforming as $(1,3)$ and the charged lepton Yukawa matrix $Y$ is a spurion transforming as $(3,\bar3)$.

Specifically, MFV means that, in Eq. (\ref{eq:dsix}),
\beq
\lambda^\prime=a\lambda+b\lambda\lambda^\dagger\lambda+{\cal O}(\lambda^5),
\eeq	
where $a$ and $b$ are numbers.  Note that, if $V_L$ and $V_R$ are the diagonalizing matrices for $\lambda$, $V_L \lambda V_R^\dagger=\lambda^{\rm diag}$, then they are also the diagonalizing matrices for $\lambda\lambda^\dagger\lambda$, $V_L\lambda\lambda^\dagger\lambda V_R^\dagger=(\lambda^{\rm diag})^3$. Then, Eqs. (\ref{eq:vlr}), (\ref{eq:deflamh}) and (\ref{eq:yneqm}) become
\beqa\label{eq:mfv}
\frac{\sqrt2m}{v} &=&\left(1+\frac{av^2}{2\Lambda^2}\right)\lambda^{\rm diag}
 +\frac{bv^2}{2\Lambda^2}(\lambda^{\rm diag})^3,\nonumber\\
\hat\lambda&=&a\lambda^{\rm diag}+b(\lambda^{\rm diag})^3
=a\frac{\sqrt2 m}{v}+\frac{2\sqrt2bm^3}{v^3},
\nonumber\\
Y_{ij}&=&\frac{\sqrt2m_i}{v}\delta_{ij}\left[1+\frac{a v^2}{\Lambda^2}
+\frac{2b m_i^2}{\Lambda^2}\right],
\eeqa
where, in the expressions for $\hat\lambda$ and $Y$, we included only the leading universal and leading non-universal corrections to the SM relations.

We learn the following points about the Higgs-related lepton flavor parameters in this class of models:
\begin{enumerate}
\item $h$ has no flavor off-diagonal couplings:
\beq\label{eq:nfcfc}
Y_{\mu\tau},Y_{\tau\mu}=0.
\eeq
\item The values of the diagonal couplings deviate from their SM values. The deviation is small, of order $v^2/\Lambda^2$:
\beq\label{eq:mfvfd}
Y_\tau\approx\left(1+\frac{av^2}{\Lambda^2}\right)\  \frac{\sqrt2 m_\tau}{v}.
\eeq
\item The ratio between the Yukawa couplings to different charged lepton flavors deviates from its SM value. The deviation is, however, very small, of order $m_\ell^2/\Lambda^2$:
\beq\label{eq:mfvr}
\frac{Y_\mu}{Y_\tau}=\frac{m_\mu}{m_\tau}\left(1-\frac{2b(m_\tau^2-m_\mu^2)}{\Lambda^2}\right).
\eeq
\end{enumerate}

The predictions of the SM with MFV non-renormalizable terms are then the following:
\beqa\label{eq:mfvh}
\left(\frac{\sigma(pp\to h)^{\rm SM}}{\sigma(pp\to h)}\frac{\Gamma_{\rm tot}}{\Gamma_{\rm tot}^{\rm SM}}\right)R_{\tau^+\tau^-}&=&1+2av^2/\Lambda^2,\nonumber\\
X_{\mu^+\mu^-}&=&(m_\mu/m_\tau)^2(1-4bm_\tau^2/\Lambda^2),\nonumber\\
X_{\tau\mu}&=&0.
\eeqa
Thus, MFV will be excluded if experiments observe the $h\to\mu\tau$ decay. On the other hand, MFV allows for a universal deviation of ${\cal O}(v^2/\Lambda^2)$ of the flavor-diagonal dilepton rates, and a smaller non-universal deviation of ${\cal O}(m_\tau^2/\Lambda^2)$.

\subsection{FN}
An attractive explanation of the smallness and hierarchy in the Yukawa couplings is provided by the Froggatt-Nielsen (FN) mechanism~\cite{Froggatt:1978nt}. In this framework, a $U(1)_H$ symmetry, under which different generations carry different charges, is broken by a small parameter $\epsilon_H$. Without loss of generality, $\epsilon_H$ is taken to be a spurion of charge $-1$. Then, various entries in the Yukawa mass matrices are suppressed by different powers of $\epsilon_H$, leading to smallness and hierarchy.

Specifically for the leptonic Yukawa matrix, taking $h$ to be neutral under $U(1)_H$, $H(h)=0$, we have
\beq
 \lambda_{ij}\propto\epsilon_H^{H(E_j)-H(L_i)}\,.
\eeq
We emphasize that the FN mechanism dictates only the parametric suppression. Each entry has an arbitrary order one coefficient.
The resulting parametric suppression of the masses and leptonic mixing angles is given by~\cite{Grossman:1995hk}
\beq
m_{\ell_i}/v\sim\epsilon_H^{H(E_i)-H(L_i)}\,,\ \ \ |U_{ij}|\sim\epsilon_H^{H(L_j)-H(L_i)}\,.
\eeq

Since $H(\phi^\dagger\phi)=0$, the entries of the matrix $\lambda^\prime$ have the same parametric suppression as the corresponding entries in $\lambda$ \cite{Leurer:1993gy}, though the order one coefficients are different:
\beq
\lambda^\prime_{ij}={\cal O}(1)\times\lambda_{ij}.
\eeq
This structure allows us to estimate the entries of $\hat\lambda_{ij}$ in terms of physical observables:
\beqa
\hat\lambda_{33}&\sim&m_\tau/v,\nonumber\\
\hat\lambda_{22}&\sim&m_\mu/v,\nonumber\\
\hat\lambda_{23}&\sim&|U_{23}|(m_\tau/v),\nonumber\\
\hat\lambda_{32}&\sim&(m_\mu/v)/|U_{23}|.
\eeqa

We learn the following points about the Higgs-related lepton flavor parameters in this class of models:
\begin{enumerate}
\item $h$ has flavor off-diagonal couplings:
\beqa\label{eq:fnfc}
Y_{\mu\tau}&=&{\cal O}\left(\frac{|U_{23}|v m_\tau}{\Lambda^2}\right),\nonumber\\
Y_{\tau\mu}&=&{\cal O}\left(\frac{v m_\mu}{|U_{23}|\Lambda^2}\right).
\eeqa
\item The values of the diagonal couplings deviate from their SM values:
\beq\label{eq:fnfd}
Y_\tau\approx\frac{\sqrt2 m_\tau}{v}\ \left[1+{\cal O}\left(\frac{v^2}{\Lambda^2}\right)\right].
\eeq
\item The ratio between the Yukawa couplings to different charged lepton flavors deviates from its SM value:
\beq\label{eq:mfvr}
\frac{Y_\mu}{Y_\tau}=\frac{m_\mu}{m_\tau}\left[1+{\cal O}\left(\frac{v^2}{\Lambda^2}\right)\right].
\eeq
\end{enumerate}

The predictions of the SM with FN-suppressed non-renormalizable terms are then the following:
\beqa\label{eq:fnh}
\left(\frac{\sigma(pp\to h)^{\rm SM}}{\sigma(pp\to h)}\frac{\Gamma_{\rm tot}}{\Gamma_{\rm tot}^{\rm SM}}\right)R_{\tau^+\tau^-}&=&1+{\cal O}(v^2/\Lambda^2),\nonumber\\
X_{\mu^+\mu^-}&=&(m_\mu/m_\tau)^2(1+{\cal O}(v^2/\Lambda^2)),\nonumber\\
X_{\tau\mu}&=&{\cal O}(v^4/\Lambda^4).
\eeqa
Thus, FN will be excluded if experiments observe deviations from the SM of the same size in both flavor-diagonal and flavor-changing $h$ decays. On the other hand, FN allows non-universal deviations of ${\cal O}(v^2/\Lambda^2)$ in the flavor-diagonal dilepton rates, and a smaller deviation of ${\cal O}(v^4/\Lambda^4)$ in the off-diagonal rate.

\section{Conclusions}
\label{sec:con}

(i) Measurements of CP violating $B$-meson decays have established
that the Kobayashi-Maskawa mechanism is the dominant source of
the observed CP violation.

(ii) Measurements of flavor changing $B$-meson decays have
established the the Cabibbo-Kobayashi-Maskawa mechanism is
a major player in flavor violation.

(iii) The consistency of all these measurements with the CKM predictions
sharpens the new physics flavor puzzle: If there is new physics at, or below,
the TeV scale, then its flavor structure must be highly non-generic.

(iv) Measurements of neutrino flavor parameters have not only not clarified
the standard model flavor puzzle, but actually deepened it. Whether they
imply an anarchical structure, or a tribimaximal mixing, it seems that
the neutrino flavor structure is very different from that of quarks.

(v) If the LHC experiments, ATLAS and CMS, discover new particles that
couple to the Standard Model fermions, then, in principle, they will
be able to measure new flavor parameters. Consequently, the new physics
flavor puzzle is likely to be understood.

(vi) If the flavor structure of such new particles is affected by the
same physics that sets the flavor structure of the Yukawa couplings, then
the LHC experiments (and future flavor factories) may be able to shed
light also on the standard model flavor puzzle.

(vii) The recently discovered Higgs-like boson provides an opportunity to make progress in our understanding of the flavor puzzle(s).

The huge progress in flavor physics in recent years has provided
answers to many questions. At the same time, new questions arise. The LHC era is likely to provide more answers and more questions.

\appendix
\section{The CKM matrix}
\label{app:ckm}
The CKM matrix $V$ is a $3\times3$ unitary matrix. Its form, however,
is not unique:

$(i)$ There is freedom in defining $V$ in that we can permute between
the various generations. This freedom is fixed by ordering the up quarks and
the down quarks by their masses, {\it i.e.} $(u_1,u_2,u_3)\to(u,c,t)$ and
$(d_1,d_2,d_3)\to(d,s,b)$. The elements of $V$ are written as follows:
\beq\label{defVij}
V=\begin{pmatrix}V_{ud}&V_{us}&V_{ub}\\
V_{cd}&V_{cs}&V_{cb}\\ V_{td}&V_{ts}&V_{tb}\\ \end{pmatrix}.
\eeq

$(ii)$ There is further freedom in the phase structure of $V$. This
means that the number of physical parameters in $V$ is smaller than
the number of parameters in a general unitary $3\times3$ matrix which
is nine (three real angles and six phases). Let us define $P_q$
($q=u,d$) to be diagonal unitary (phase) matrices. Then, if instead of
using $V_{qL}$ and $V_{qR}$ for the rotation (\ref{diagMq}) to the
mass basis we use $\tilde V_{qL}$ and $\tilde V_{qR}$, defined by
$\tilde V_{qL}=P_q V_{qL}$ and $\tilde V_{qR}=P_q V_{qR}$, we still
maintain a legitimate mass basis since $M_q^{\rm diag}$ remains
unchanged by such transformations. However, $V$ does change:
\beq\label{eqphase}
V\to P_u V P_d^*.
\eeq
This freedom is fixed by demanding that $V$ has the minimal number of
phases. In the three generation case $V$ has a single phase. (There
are five phase differences between the elements of $P_u$ and $P_d$ and,
therefore, five of the six phases in the CKM matrix can be removed.) This is
the Kobayashi-Maskawa phase $\delta_{\rm KM}$ which is the single source of
CP violation in the quark sector of the Standard Model \cite{Kobayashi:1973fv}.

The fact that $V$ is unitary and depends on only four independent
physical parameters can be made manifest by choosing a specific
parametrization. The standard choice is \cite{Chau:1984fp}
\beq\label{stapar}
V=\begin{pmatrix}c_{12}c_{13}&s_{12}c_{13}&
s_{13}e^{-i\delta}\\
-s_{12}c_{23}-c_{12}s_{23}s_{13}e^{i\delta}&
c_{12}c_{23}-s_{12}s_{23}s_{13}e^{i\delta}&s_{23}c_{13}\\
s_{12}s_{23}-c_{12}c_{23}s_{13}e^{i\delta}&
-c_{12}s_{23}-s_{12}c_{23}s_{13}e^{i\delta}&c_{23}c_{13}\\ \end{pmatrix},
\eeq
where $c_{ij}\equiv\cos\theta_{ij}$ and $s_{ij}\equiv\sin\theta_{ij}$.
The $\theta_{ij}$'s are the three real mixing parameters while
$\delta$ is the Kobayashi-Maskawa phase. It is known experimentally
that $s_{13}\ll s_{23}\ll s_{12}\ll1$. It is convenient to choose an
approximate expression where this hierarchy is manifest. This is the
Wolfenstein parametrization, where the four mixing parameters are
$(\lambda,A,\rho,\eta)$ with $\lambda=|V_{us}|=0.23$ playing the role
of an expansion parameter and $\eta$ representing the CP violating
phase \cite{Wolfenstein:1983yz,Buras:1994ec}:
\beq\label{wolpar}
V=\begin{pmatrix}
1-\frac12\lambda^2-\frac18\lambda^4 & \lambda &
A\lambda^3(\rho-i\eta)\\
-\lambda +\frac12A^2\lambda^5[1-2(\rho+i\eta)] &
1-\frac12\lambda^2-\frac18\lambda^4(1+4A^2) & A\lambda^2 \\
A\lambda^3[1-(1-\frac12\lambda^2)(\rho+i\eta)]&
-A\lambda^2+\frac12A\lambda^4[1-2(\rho+i\eta)]
& 1-\frac12A^2\lambda^4 \\ \end{pmatrix}\; .
\eeq

\begin{figure}[tb]
  \centering
  {\includegraphics[width=0.65\textwidth]{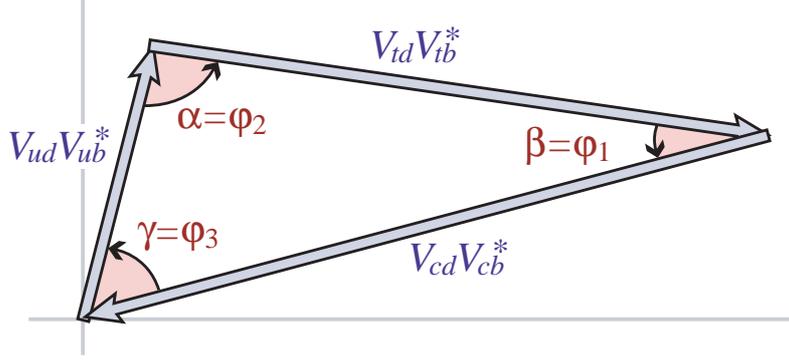}}
  \caption{Graphical representation of the unitarity constraint
  $V_{ud}V_{ub}^*+V_{cd}V_{cb}^*+V_{td}V_{tb}^*=0$ as a triangle in
  the complex plane.}
  \label{fg:tri}
\end{figure}

A very useful concept is that of the {\it unitarity triangles}. The
unitarity of the CKM matrix leads to various relations among the
matrix elements, {\it e.g.}
\beqa\label{Unitds}
V_{ud}V_{us}^*+V_{cd}V_{cs}^*+V_{td}V_{ts}^*=0,\\
\label{Unitsb}
V_{us}V_{ub}^*+V_{cs}V_{cb}^*+V_{ts}V_{tb}^*=0,\\
\label{Unitdb}
V_{ud}V_{ub}^*+V_{cd}V_{cb}^*+V_{td}V_{tb}^*=0.
\eeqa
Each of these three relations requires
the sum of three complex quantities to vanish and so can be geometrically
represented in the complex plane as a triangle. These are
``the unitarity triangles", though the term ``unitarity triangle"
is usually reserved for the relation (\ref{Unitdb}) only. The
unitarity triangle related to Eq. (\ref{Unitdb}) is depicted in
Fig. \ref{fg:tri}.

The rescaled unitarity triangle  is derived from (\ref{Unitdb})
by (a) choosing a phase convention such that $(V_{cd}V_{cb}^*)$
is real, and (b) dividing the lengths of all sides by $|V_{cd}V_{cb}^*|$.
Step (a) aligns one side of the triangle with the real axis, and
step (b) makes the length of this side 1. The form of the triangle
is unchanged. Two vertices of the rescaled unitarity triangle are
thus fixed at (0,0) and (1,0). The coordinates of the remaining
vertex correspond to the Wolfenstein parameters $(\rho,\eta)$.
The area of the rescaled unitarity triangle is $|\eta|/2$.

Depicting the rescaled unitarity triangle in the
$(\rho,\eta)$ plane, the lengths of the two complex sides are
\beq\label{RbRt}
R_u\equiv\left|{V_{ud}V_{ub}\over V_{cd}V_{cb}}\right|
=\sqrt{\rho^2+\eta^2},\ \ \
R_t\equiv\left|{V_{td}V_{tb}\over V_{cd}V_{cb}}\right|
=\sqrt{(1-\rho)^2+\eta^2}.
\eeq
The three angles of the unitarity triangle are defined as follows
\cite{Dib:1989uz,Rosner:1988nx}:
\beq\label{abcangles}
\alpha\equiv\arg\left[-{V_{td}V_{tb}^*\over V_{ud}V_{ub}^*}\right],\ \ \
\beta\equiv\arg\left[-{V_{cd}V_{cb}^*\over V_{td}V_{tb}^*}\right],\ \ \
\gamma\equiv\arg\left[-{V_{ud}V_{ub}^*\over V_{cd}V_{cb}^*}\right].
\eeq
They are physical quantities and can be independently measured by CP
asymmetries in $B$ decays. It is also useful to define the two
small angles of the unitarity triangles (\ref{Unitsb},\ref{Unitds}):
\beq\label{bbangles}
\beta_s\equiv\arg\left[-{V_{ts}V_{tb}^*\over V_{cs}V_{cb}^*}\right],\ \ \
\beta_K\equiv\arg\left[-{V_{cs}V_{cd}^*\over V_{us}V_{ud}^*}\right].
\eeq

\section{CPV in $B$ decays to final CP eigenstates}
\label{sec:formalism}
We define decay amplitudes of $B$ (which could be charged or neutral)
and its CP conjugate $\Bbar$ to a multi-particle final state $f$ and its
CP conjugate $\fb$ as
\beq\label{decamp}
A_{\f}=\langle \f|{\cal H}|B\rangle\quad , \quad
\overline{A}_{\f}=\langle \f|{\cal H}|\Bbar\rangle\quad , \quad
A_{\fb}=\langle \fb|{\cal H}|B\rangle\quad , \quad
\overline{A}_{\fb}=\langle \fb|{\cal H}|\Bbar\rangle\; ,
\eeq
where ${\cal H}$ is the Hamiltonian governing
weak interactions.  The action of CP on these states introduces
phases $\xi_B$ and $\xi_f$ according to
\beqa\label{eq:phaseconv}
\CP|B\rangle &=& e^{+i\xi_{B}}\,|\Bbar\rangle \quad , \quad
\CP|\f\rangle = e^{+i\xi_{\f}}\,|\fb\rangle \; ,\no\\
\CP|\Bbar\rangle& =& e^{-i\xi_{B}}\,|B\rangle \quad , \quad
\CP|\fb\rangle = e^{-i\xi_{\f}}\,|\f\rangle \ ,
\eeqa
so that $(\CP)^2=1$. The phases $\xi_B$ and $\xi_f$ are arbitrary and
unphysical because of the flavor symmetry of the strong
interaction. If CP is conserved by the dynamics, $[\CP,{\cal H}] =
0$, then $A_f$ and $\overline{A}_{\fb}$ have the same magnitude and an
arbitrary unphysical relative phase
\beq\label{spupha}
\overline{A}_{\fb} = e^{i(\xi_{\f}-\xi_{B})}\, A_f\; .
\eeq

A state that is initially a superposition of $\Bz$ and $\Bzb$, say
\beq
|\psi(0)\rangle = a(0)|\Bz\rangle+b(0)|\Bzb\rangle \; ,
\eeq
will evolve in time acquiring components that describe all possible
decay final states $\{f_1,f_2,\ldots\}$, that is,
\beq
|\psi(t)\rangle =
a(t)|\Bz\rangle+b(t)|\Bzb\rangle+c_1(t)|f_1\rangle+c_2(t)|f_2\rangle+\cdots
\; .
\eeq
If we are interested in computing only the values of $a(t)$ and $b(t)$
(and not the values of all $c_i(t)$), and if the times $t$ in which we
are interested are much larger than the typical strong interaction
scale, then we can use a much simplified
formalism~\cite{Weisskopf:au}. The simplified time evolution is
determined by a $2\times 2$ effective Hamiltonian $\Heff$ that is
not Hermitian, since otherwise the mesons would only oscillate and not
decay. Any complex matrix, such as $\Heff$, can be written in terms of
Hermitian matrices $\Meff$ and $\Geff$ as
\beq
\Heff = \Meff - \frac{i}{2}\,\Geff \; .
\eeq
$\Meff$ and $\Geff$ are associated with
$(\Bz,\Bzb)\leftrightarrow(\Bz,\Bzb)$ transitions via off-shell
(dispersive) and on-shell (absorptive) intermediate states, respectively.
Diagonal elements of $\Meff$ and $\Geff$ are associated with the
flavor-conserving transitions $\Bz\to\Bz$ and $\Bzb\to\Bzb$ while
off-diagonal elements are associated with flavor-changing transitions
$\Bz\leftrightarrow\Bzb$.

The eigenvectors of $\Heff$ have well defined masses and decay
widths. We introduce complex parameters $p$ and $q$ to
specify the components of the strong interaction eigenstates, $\Bz$ and
$\Bzb$, in the light ($B_L$) and heavy ($B_H$) mass eigenstates:
\beq\label{defpq}
|B_{L,H}\rangle=p|\Bz\rangle\pm q|\Bzb\rangle
\eeq
with the normalization $|p|^2+|q|^2=1$. The special form of Eq. (\ref{defpq}) is related to the fact that CPT imposes $\Meff_{11} = \Meff_{22}$ and $\Geff_{11}=\Geff_{22}$. Solving the eigenvalue problem gives
\beq\label{eq:qpgm}
\left(\frac{q}{p}\right)^2=\frac{\Meff_{12}^\ast -
    (i/2)\Geff_{12}^\ast}{\Meff_{12}-(i/2)\Geff_{12}}\; .
\eeq
If either CP or T is a symmetry of $\Heff$, then $\Meff_{12}$ and $\Geff_{12}$ are
relatively real, leading to
\beq
\left(\frac{q}{p}\right)^2 = e^{2i\xi_B} \quad \Rightarrow \quad
\left|\frac{q}{p}\right| = 1 \; ,
\eeq
where $\xi_B$ is the arbitrary unphysical phase introduced in
Eq.~(\ref{eq:phaseconv}).

The real and imaginary parts of the eigenvalues of $\Heff$
corresponding to $|B_{L,H}\rangle$ represent their masses and
decay-widths, respectively. The mass difference $\Delta m_B$ and the
width difference $\Delta\Gamma_B$ are defined as follows:
\beq\label{DelmG}
\Delta m_B\equiv M_H-M_L,\ \ \ \Delta\Gamma_B\equiv\Gamma_H-\Gamma_L.
\eeq
Note that here $\Delta m_B$ is positive by definition, while the sign of
$\Delta\Gamma_B$ is to be experimentally determined.
The average mass and width are given by
\beq\label{aveMG}
m_B\equiv{M_H+M_L\over2},\ \ \ \Gamma_B\equiv{\Gamma_H+\Gamma_L\over2}.
\eeq
It is useful to define dimensionless ratios $x$ and $y$:
\beq\label{defxy}
x\equiv{\Delta m_B\over\Gamma_B},\ \ \ y\equiv{\Delta\Gamma_B\over2\Gamma_B}.
\eeq
Solving the eigenvalue equation gives
\beq\label{eveq}
(\Delta m_B)^2-{1\over4}(\Delta\Gamma_B)^2=(4|M_{12}|^2-|\Gamma_{12}|^2),\ \ \ \
\Delta m_B\Delta\Gamma_B=4\re{M_{12}\Gamma_{12}^*}.
\eeq

All CP-violating observables in $B$ and $\Bbar$ decays to final states $f$
and $\fb$ can be expressed in terms of phase-convention-independent
combinations of $A_f$, $\overline{A}_f$, $A_{\overline{f}}$ and
$\overline{A}_{\overline{f}}$, together with, for neutral-meson decays
only, $q/p$. CP violation in charged-meson decays depends only on the
combination $|\overline{A}_{\fb}/A_f|$, while CP violation in
neutral-meson decays is complicated by $\Bz\leftrightarrow\Bzb$
oscillations and depends, additionally, on $|q/p|$ and on $\lambda_f
\equiv (q/p)(\overline{A}_f/A_f)$.

For neutral $D$, $B$, and $B_s$ mesons, $\Delta\Gamma/\Gamma\ll1$ and
so both mass eigenstates must be considered in their evolution. We
denote the state of an initially pure $|\Bz\rangle$ or $|\Bzb\rangle$
after an elapsed proper time $t$ as $|\Bz_{\mathrm{phys}}(t)\rangle$
or $|\Bzb_{\mathrm{phys}}(t)\rangle$, respectively. Using the
effective Hamiltonian approximation, we obtain
\beqa\label{defphys}
|\Bz_{\rm phys}(t)\rangle&=&g_+(t)\,|\Bz\rangle
- \frac qp\ g_-(t)|\Bzb\rangle,\no\\
|\Bzb_{\rm phys}(t)\rangle&=&g_+(t)\,|\Bzb\rangle
- \frac pq\ g_-(t)|\Bz\rangle \; ,
\eeqa
where
\beq
g_\pm(t) \equiv \frac12\left(e^{-im_Ht-\frac12\Gamma_Ht}\pm
  e^{-im_Lt-\frac12\Gamma_Lt}\right).
\eeq

One obtains the following time-dependent decay rates:
\beqa
\frac{d\Gamma[\Bz_{\rm phys}(t)\to f]/dt}{e^{-\Gamma t}{\cal N}_f}&=&
\left(|A_f|^2+|(q/p)\overline{A}_f|^2\right)\cosh(y\Gamma t)
  +\left(|A_f|^2-|(q/p)\overline{A}_f|^2\right)\cos(x\Gamma t)\no\\
&+&2\,\re{(q/p)A_f^\ast \overline{A}_f}\sinh(y\Gamma t)
-2\,\im{(q/p)A_f^\ast \overline{A}_f}\sin(x\Gamma t)
\label{decratbt1}\;,\\
\frac{d\Gamma[\Bzb_{\rm phys}(t)\to f]/dt}{e^{-\Gamma t}{\cal N}_f}&=&
\left(|(p/q)A_f|^2+|\overline{A}_f|^2\right)\cosh(y\Gamma t)
  -\left(|(p/q)A_f|^2-|\overline{A}_f|^2\right)\cos(x\Gamma t)\no\\
&+&2\,\re{(p/q)A_f\overline{A}^\ast_f}\sinh(y\Gamma t)
-2\,\im{(p/q)A_f\overline{A}^\ast_f}\sin(x\Gamma t)
\label{decratbt2}\; ,
\eeqa
where ${\cal N}_f$ is a common normalization factor. Decay rates to
the CP-conjugate final state $\fb$ are obtained analogously, with
${\cal N}_f = {\cal N}_{\fb}$ and the substitutions $A_f\to A_{\fb}$
and $\overline{A}_f\to\overline{A}_{\fb}$ in
Eqs.~(\ref{decratbt1},\ref{decratbt2}). Terms proportional
to $|A_f|^2$ or $|\overline{A}_f|^2 $ are associated with decays that
occur without any net $B\leftrightarrow\Bbar$ oscillation, while terms
proportional to $|(q/p)\overline{A}_f|^2$ or $|(p/q)A_f|^2$ are
associated with decays following a net oscillation. The $\sinh(y\Gamma
t)$ and $\sin(x\Gamma t)$ terms of
Eqs.~(\ref{decratbt1},\ref{decratbt2}) are associated with the
interference between these two cases. Note that, in multi-body decays,
amplitudes are functions of phase-space variables. Interference may
be present in some regions but not others, and is strongly influenced
by resonant substructure.

One possible manifestation of CP-violating effects in meson decays
\cite{Nir:1992uv} is in the interference between a decay without
mixing, $\Bz\to f$, and a decay with mixing, $\Bz\to \Bzb\to f$ (such
an effect occurs only in decays to final states that are common to
$\Bz$ and $\Bzb$, including all CP eigenstates). It is defined by
\beq\label{cpvint}
\im{\lambda_f}\ne 0 \; ,
\eeq
with
\beq\label{deflam}
\lambda_f \equiv \frac{q}{p}\frac{\overline{A}_f}{A_f} \; .
\eeq
This form of CP violation can be observed, for example, using the
asymmetry of neutral meson decays into final CP eigenstates $f_{\CP}$
\beq\label{asyfcp}
{\cal A}_{f_{\CP}}(t)\equiv\frac{d\Gamma/dt[\Bzb_{\rm phys}(t)\to f_{\CP}]-
d\Gamma/dt[\Bz_{\rm phys}(t)\to f_{\CP}]}
{d\Gamma/dt[\Bzb_{\rm phys}(t)\to f_{\CP}]+d\Gamma/dt[\Bz_{\rm phys}(t)\to
  f_{\CP}]}\; .
\eeq
For $\Delta\Gamma = 0$ and $|q/p|=1$ (which is a good approximation
for $B$ mesons), ${\cal A}_{f_{\CP}}$ has a particularly simple form
\cite{Dunietz:1986vi,Blinov:ru,Bigi:1986vr}:
\beqa\label{asyfcpb}
{\cal A}_{f}(t)&=&S_f\sin(\Delta mt)-C_f\cos(\Delta mt),\no\\
S_f&\equiv&\frac{2\,\im{\lambda_{f}}}{1+|\lambda_{f}|^2},\ \ \
C_f\equiv\frac{1-|\lambda_{f}|^2}{1+|\lambda_{f}|^2}
\; ,
\eeqa

Consider the $B\to f$ decay amplitude $A_f$, and the CP conjugate
process, $\Bbar\to\fb$, with decay amplitude $\overline{A}_{\fb}$. There
are two types of phases that may appear in these decay amplitudes.
Complex parameters in any Lagrangian term that contributes to the
amplitude will appear in complex conjugate form in the CP-conjugate
amplitude. Thus their phases appear in $A_f$ and
$\overline{A}_{\overline{f}}$ with opposite signs. In the Standard
Model, these phases occur only in the couplings of the $W^\pm$ bosons
and hence are often called ``weak phases''. The weak phase of any
single term is convention dependent. However, the difference between
the weak phases in two different terms in $A_f$ is convention
independent. A second type of phase can appear in scattering or decay
amplitudes even when the Lagrangian is real. Their origin is the
possible contribution from intermediate on-shell states in the decay
process. Since these phases are generated by CP-invariant
interactions, they are the same in $A_f$ and
$\overline{A}_{\overline{f}}$. Usually the dominant rescattering is
due to strong interactions and hence the designation ``strong phases''
for the phase shifts so induced. Again, only the relative strong
phases between different terms in the amplitude are physically
meaningful.

The `weak' and `strong' phases discussed here appear in addition to
the `spurious' CP-transformation phases of Eq.~(\ref{spupha}). Those
spurious phases are due to an arbitrary choice of phase
convention, and do not originate from any dynamics or induce any \CP
violation. For simplicity, we set them to zero from here on.

It is useful to write each contribution $a_i$ to $A_f$ in three parts:
its magnitude $|a_i|$, its weak phase $\phi_i$, and its strong
phase $\delta_i$. If, for example, there are two such
contributions, $A_f = a_1 + a_2$, we have
\beqa\label{weastr}
A_f&=& |a_1|e^{i(\delta_1+\phi_1)}+|a_2|e^{i(\delta_2+\phi_2)},\no\\
\overline{A}_{\overline{f}}&=&
|a_1|e^{i(\delta_1-\phi_1)}+|a_2|e^{i(\delta_2-\phi_2)}.
\eeqa
Similarly, for neutral meson decays, it is useful to write
\beq\label{defmgam}
\Meff_{12} = |\Meff_{12}| e^{i\phi_M} \quad , \quad
\Geff_{12} = |\Geff_{12}| e^{i\phi_\Gamma} \; .
\eeq
Each of the phases appearing in Eqs.~(\ref{weastr},\ref{defmgam}) is
convention dependent, but combinations such as $\delta_1-\delta_2$,
$\phi_1-\phi_2$, $\phi_M-\phi_\Gamma$ and $\phi_M+\phi_1-\overline{\phi}_1$
(where $\overline{\phi}_1$ is a weak phase contributing to $\overline{A}_f$)
are physical.

In the approximations that only a single weak phase contributes to decay,
$A_f=|a_f|e^{i(\delta_f+\phi_f)}$, and that
$|\Geff_{12}/\Meff_{12}|=0$, we obtain $|\lambda_f|=1$ and
the \CP asymmetries in decays to a final CP
eigenstate $f$ [Eq. (\ref{asyfcp})] with eigenvalue $\eta_f= \pm 1$
are given by
\beq\label{afcth}
{\cal A}_{f_{\CP}}(t) = \im{\lambda_f}\; \sin(\Delta m t) \; \
\mathrm{with}\ \
\im{\lambda_f}=\eta_f\sin(\phi_M+2\phi_f).
\eeq
Note that the phase so measured is purely a weak phase, and no
hadronic parameters are involved in the extraction of its value from
$\im{\lambda_f}$.

\section{Supersymmetric flavor violation}
\label{app:susyfv}
%
\subsection{Mass insertions}
\label{app:susydel}
Supersymmetric models provide, in general, new sources of flavor
violation. We here present the formalism of mass insertions. We do
that for the charged sleptons, but the formalism is straightforwardly
adapted for squarks.

The supersymmetric lepton flavor violation is most commonly analyzed
in the basis in which the charged lepton mass matrix and the gaugino
vertices are diagonal. In this basis, the slepton masses are not
necessarily flavor-diagonal, and have the form
\beq
\widetilde\ell_{Mi}^*(M^2_{\widetilde\ell})^{MN}_{ij}\widetilde\ell_{Nj}=
(\widetilde\ell_{Li}^*\ \widetilde\ell_{Rk}^*)
\left(\begin{array}{cc}
    M^2_{Lij} & A_{il}v_d \\ A_{jk}v_d & M^2_{Rkl}\end{array}\right)
\left(\begin{array}{c} \widetilde\ell_{Lj}\\
    \widetilde\ell_{Rl}\end{array}\right),
\eeq
where $M,N=L,R$ label chirality, and $i,j,k,l=1,2,3$ are generational
indices. $M^2_L$ and $M^2_R$ are the supersymmetry breaking slepton
masses-squared. The $A$ parameters enter in the trilinear scalar
couplings $A_{ij}\phi_d\widetilde\ell_{Li}\widetilde\ell_{Rj}^*$,
where $\phi_d$ is the down-type Higgs boson, and
$v_d=\langle\phi_d\rangle$. We neglect small flavor-conserving terms
involving $\tan\beta=v_u/v_d$.

In this basis, charged LFV takes place through one or more slepton
mass insertion. Each mass insertion brings with it a factor of
\beq\label{eq:defdeli}
\delta^{MN}_{ij}\equiv(M^2_{\widetilde\ell})^{MN}_{ij}/\tilde m^2,
\eeq
where $\tilde m^2$ is the representative slepton mass scale. Physical
processes therefore constrain
\beq
(\delta^{MN}_{ij})_{\rm eff}\sim{\rm max}\left[
  \delta^{MN}_{ij},\delta^{MP}_{ik}\delta^{PN}_{kj},\ldots,
  (i\leftrightarrow j)\right].
\eeq
For example,
\beq
(\delta^{LR}_{12})_{\rm eff}\sim{\rm max}\left[
  A_{12}v_d/\tilde m^2,M^2_{L1k}A_{k2}v_d/\tilde
  m^4,A_{1k}v_dM^2_{Rk2}/\tilde m^4,\ldots,(1\leftrightarrow2)\right].
\eeq
Note that contributions with two or more insertions may be less
suppressed than those with only one.

It is useful to express the $\delta^{MN}_{ij}$ mass insertions in
terms of parameters in the mass basis. We can write, for example,
\beq
\delta^{LL}_{ij}=\frac{1}{\tilde m^2}\sum_\alpha
K^L_{i\alpha}K^{L*}_{j\alpha}\Delta\tilde m^2_{L\alpha}.
\eeq
Here, we ignore $L-R$ mixing, so that $K^L_{i\alpha}$ is the mixing
angle in the coupling of a neutralino to
$\ell_{Li}-\widetilde\ell_{L\alpha}$ (with $\ell_i=e,\mu,\tau$
denoting charged lepton mass eigenstates and
$\widetilde\ell_\alpha=\widetilde\ell_1,\widetilde\ell_2,\widetilde\ell_3$
denoting charged slepton mass eigenstates), and $\Delta\tilde
m^2_{L\alpha}=m^2_{\widetilde\ell_{L\alpha}}-\tilde m^2$. Using the
unitarity of the mixing matrix $K^L$, we can write
\beq
\tilde m^2\delta^{LL}_{ij}=\sum_\alpha K^L_{i\alpha}K^{L*}_{j\alpha}
(\Delta\tilde m^2_{L\alpha}+\tilde m^2)=(M^2_{\widetilde\ell})^{LL}_{ij},
\eeq
thus reproducing the definition (\ref{eq:defdeli}).

In many cases, a two generation effective framework is useful. To
understand that, consider a case where (no summation over $i,j,k$)
\beqa
|K^L_{ik}K^{L*}_{jk}|&\ll&|K^L_{ij}K^{L*}_{j}|,\no\\
|K^L_{ik}K^{L*}_{jk}\Delta m^2_{\widetilde\ell_{Lk}\widetilde\ell_{Li}}|
&\ll&|K^L_{ij}K^{L*}_{j}\Delta
m^2_{\widetilde\ell_{Lj}\widetilde\ell_{Li}}|,
\eeqa
where $\Delta m^2_{\widetilde\ell_j\widetilde\ell_i}=
m^2_{\widetilde\ell_{Lj}}-m^2_{\widetilde\ell_{Li}}$. Then, the
contribution of the intermediate $\widetilde\ell_k$ can be neglected
and, furthermore, to a good approximation
$K^L_{ii}K^{L*}_{ji}+K^L_{ij}K^{L*}_{jj}=0$. For these cases, we
obtain
\beq
\delta^{LL}_{ij}=\frac{\Delta m^2_{\widetilde\ell_{Lj}\widetilde\ell_{Li}}}
{\tilde m^2}K^L_{ij}K^{L*}_{jj}.
\eeq
%

\subsection{Neutral meson mixing}
\label{app:susyd}
We consider the squark-gluino box diagram contribution to
$D^0-\overline{D}^0$ mixing amplitude that is proportional to
$K_{2i}^u K^{u*}_{1i}K_{2j}^u K^{u*}_{1j}$, where $K^u$ is the
mixing matrix of the gluino couplings to left-handed up quarks and
their up squark partners. (In the language of the mass insertion
approximation, we calculate here the contribution that is $\propto
[(\delta^u_{LL})_{12}]^2$.) We work in the mass basis for both quarks
and squarks.

The contribution is given by
\beq\label{motsusy}
M_{12}^D=-i\frac{4\pi^2}{27}\alpha_s^2m_Df_D^2B_D\eta_{\rm QCD}
\sum_{i,j}(K_{2i}^uK_{1i}^{u*}K_{2j}^uK_{1j}^{u*})(11\tilde
I_{4ij}+4\tilde m_g^2I_{4ij}).
\eeq
where
\beqa
\tilde I_{4ij}&\equiv&\int\frac{d^4p}{(2\pi)^4}\frac{p^2}{(p^2-\tilde
  m_g^2)^2(p^2-\tilde m_i^2)(p^2-\tilde m_j^2)}\no\\
&=&\frac{i}{(4\pi)^2}\left[\frac{\tilde m_g^2}
  {(\tilde m_i^2-\tilde m_g^2)(\tilde m_j^2-\tilde m_g^2)}\right.\no\\
   && +\left.\frac{\tilde m_i^4}
  {(\tilde m_i^2-\tilde m_j^2)(\tilde m_i^2-\tilde
    m_g^2)^2}\ln\frac{\tilde m_i^2}{\tilde m_g^2}
  +\frac{\tilde m_j^4}
  {(\tilde m_j^2-\tilde m_i^2)(\tilde m_j^2-\tilde
    m_g^2)^2}\ln\frac{\tilde m_j^2}{\tilde m_g^2}\right],
\eeqa
\beqa
I_{4ij}&\equiv&\int\frac{d^4p}{(2\pi)^4}\frac{1}{(p^2-\tilde
  m_g^2)^2(p^2-\tilde m_i^2)(p^2-\tilde m_j^2)}\no\\
&=&\frac{i}{(4\pi)^2}\left[\frac{1}
  {(\tilde m_i^2-\tilde m_g^2)(\tilde m_j^2-\tilde m_g^2)}\right.\no\\
   && +\left.\frac{\tilde m_i^2}
  {(\tilde m_i^2-\tilde m_j^2)(\tilde m_i^2-\tilde
    m_g^2)^2}\ln\frac{\tilde m_i^2}{\tilde m_g^2}
  +\frac{\tilde m_j^2}
  {(\tilde m_j^2-\tilde m_i^2)(\tilde m_j^2-\tilde
    m_g^2)^2}\ln\frac{\tilde m_j^2}{\tilde m_g^2}\right].
\eeqa

We now follow the discussion in refs. \cite{Raz:2002zx,Nir:2002ah}.
To see the consequences of the super-GIM mechanism, let us expand the
expression for the box integral around some value $\tilde m^2_q$ for
the squark masses-squared:
\beqa
I_4(\tilde m_g^2,\tilde m_i^2,\tilde m_j^2)&=&
I_4(\tilde m_g^2,\tilde m_q^2+\delta\tilde m_i^2,\tilde
m_q^2+\delta\tilde m_j^2)\no\\
&=&I_4(\tilde m_g^2,\tilde m_q^2,\tilde m_q^2)
+(\delta\tilde m_i^2+\delta\tilde m_j^2)I_5(\tilde m_g^2,\tilde
m_q^2,\tilde m_q^2,\tilde m_q^2)\no\\
&+&\frac12\left[(\delta\tilde m_i^2)^2+(\delta\tilde
  m_j^2)^2+2(\delta\tilde m_i^2)(\delta\tilde m_j^2)\right]I_6(\tilde m_g^2,\tilde
m_q^2,\tilde m_q^2,\tilde m_q^2,\tilde m_q^2)+\cdots,
\eeqa
where
\beq
I_n(\tilde m_g^2,\tilde m_q^2,\ldots,\tilde
m_q^2)\equiv\int\frac{d^4p}{(2\pi)^4}\frac{1}{(p^2-\tilde
  m_g^2)^2(p^2-\tilde m_q^2)^{n-2}},
\eeq
and similarly for $\tilde I_{4ij}$. Note that $I_n\propto(\tilde
m_q^2)^{n-2}$ and $\tilde I_n\propto(\tilde m_q^2)^{n-3}$. Thus, using
$x\equiv\tilde m_g^2/\tilde m_q^2$, it is customary to define
\beq
I_n\equiv\frac{i}{(4\pi)^2(\tilde m_q^2)^{n-2}}f_n(x),\ \ \ \
\tilde I_n\equiv\frac{i}{(4\pi)^2(\tilde m_q^2)^{n-3}}\tilde f_n(x).
\eeq
The unitarity of the mixing matrix implies that
\beq
\sum_i (K_{2i}^uK_{1i}^{u*}K_{2j}^uK_{1j}^{u*})=
\sum_j (K_{2i}^uK_{1i}^{u*}K_{2j}^uK_{1j}^{u*})=0.
\eeq
Consequently, the terms that are proportional $f_4,\tilde f_4,f_5$ and
$\tilde f_5$ vanish in their contribution to $M_{12}$. When
$\delta\tilde m_i^2\ll\tilde m_q^2$ for all $i$, the
leading contributions to $M_{12}$ come from $f_6$ and $\tilde f_6$. We
learn that for quasi-degenerate squarks, the leading contribution is
quadratic in the small mass-squared difference. The functions $f_6(x)$
and $\tilde f_6(x)$ are given by
\beqa
f_6(x)&=&\frac{6(1+3x)\ln x+x^3-9x^2-9x+17}{6(1-x)^5},\no\\
\tilde f_6(x)&=&\frac{6x(1+x)\ln x-x^3-9x^2+9x+1}{3(1-x)^5}.
\eeqa
For example, with $x=1$, $f_6(1)=-1/20$ and $\tilde f_6=+1/30$;
with $x=2.33$, $f_6(2.33)=-0.015$ and $\tilde f_6=+0.013$.

To further simplify things, let us consider a two generation
case. Then
\beqa
M_{12}^D&\propto& 2(K_{21}^uK_{11}^{u*})^2(\delta\tilde
m_1^2)^2+2(K_{22}^uK_{12}^{u*})^2(\delta\tilde
m_2^2)^2+(K_{21}^uK_{11}^{u*}K_{22}^uK_{12}^{u*})(\delta\tilde
m_1^2+\delta\tilde m_2^2)^2\no\\
&=&(K^u_{21}K_{11}^{u*})^2(\tilde m_2^2-\tilde m_1^2)^2.
\eeqa
We thus rewrite Eq.~(\ref{motsusy}) for the case of quasi-degenerate squarks:
\beq\label{motsusyd}
M_{12}^D=\frac{\alpha_s^2m_Df_D^2B_D\eta_{\rm QCD}}{108\tilde m_q^2}
[11\tilde f_6(x)+4xf_6(x)]\frac{(\Delta\tilde m^2_{21})^2}{\tilde m_q^4}
(K_{21}^uK_{11}^{u*})^2.
\eeq
For example, for $x=1$, $11\tilde f_6(x)+4xf_6(x)=+0.17$.
For $x=2.33$, $11\tilde f_6(x)+4xf_6(x)=+0.003$.

\section*{Acknowledgements}

I thank my students --  Yonit Hochberg, Daniel Grossman, Aielet Efrati and Avital Dery -- for many useful discussions.
The research of Y.N. is supported by the I-CORE Program of the Planning and Budgeting Committee and the Israel Science Foundation (grant No 1937/12), by the Israel Science Foundation (grant No 579/11), and by the German--Israeli Foundation (GIF) (Grant No G-1047-92.7/2009).

\section*{Bibliography}

G.C. Branco, L. Lavoura and J.P. Silva, \emph{CP Violation}
(Oxford University Press, Oxford, 1999).\\
H.R. Quinn and Y. Nir, \emph{The Mystery of the Missing Antimatter}
(Princeton University Press, Princeton, 2007).

\end{document}